 \documentclass[final,5p,times,twocolumn,authoryear]{elsarticle}

\usepackage{amssymb, amsmath, amsfonts}
\usepackage{multicol, multirow, tabularx, makecell}
\usepackage{algorithmic, graphicx, textcomp}
\usepackage{lipsum}
\usepackage{eucal, booktabs}
\usepackage{bm}
\usepackage{xcolor}
\usepackage{svg}
\usepackage{subcaption}

\journal{Astronomy $\&$ Computing}

\begin{document}

\begin{frontmatter}



\title{A Survey of Deep Learning for Complex Speech Spectrograms}


\author[first]{Yuying Xie}
\author[second]{Zheng-Hua Tan}
\affiliation[first, second]{
            organization={Department of Electronic Systems}, 
            city={Aalborg University},
            postcode={9220}, 
            country={Denmark}}

\begin{abstract}

Recent advancements in deep learning have significantly impacted the field of speech signal processing, particularly in the analysis and manipulation of complex spectrograms. 
This survey provides a comprehensive overview of the state-of-the-art techniques leveraging deep neural networks for processing complex spectrograms, which encapsulate both magnitude and phase information. 
We begin by introducing complex spectrograms and their associated features for various speech processing tasks. 
Next, we examine the key components and architectures of complex-valued neural networks, which are specifically designed to handle complex-valued data and have been applied to complex spectrogram processing. 
As recent studies have primarily focused on applying real-valued neural networks to complex spectrograms, we revisit these approaches and their architectural designs.
We then discuss various training strategies and loss functions tailored for training neural networks to process and model complex spectrograms. 
The survey further examines key applications, including phase retrieval, speech enhancement, and speaker separation, where deep learning has achieved significant progress by leveraging complex spectrograms or their derived feature representations.
Additionally, we examine the intersection of complex spectrograms with generative models. 
This survey aims to serve as a valuable resource for researchers and practitioners in the field of speech signal processing, deep learning and related fields.
\end{abstract}


\end{frontmatter}


\section{Introduction}
\label{introduction}

\begin{figure*}[htbp]
    \centering
    \includegraphics[width=0.65\textwidth]{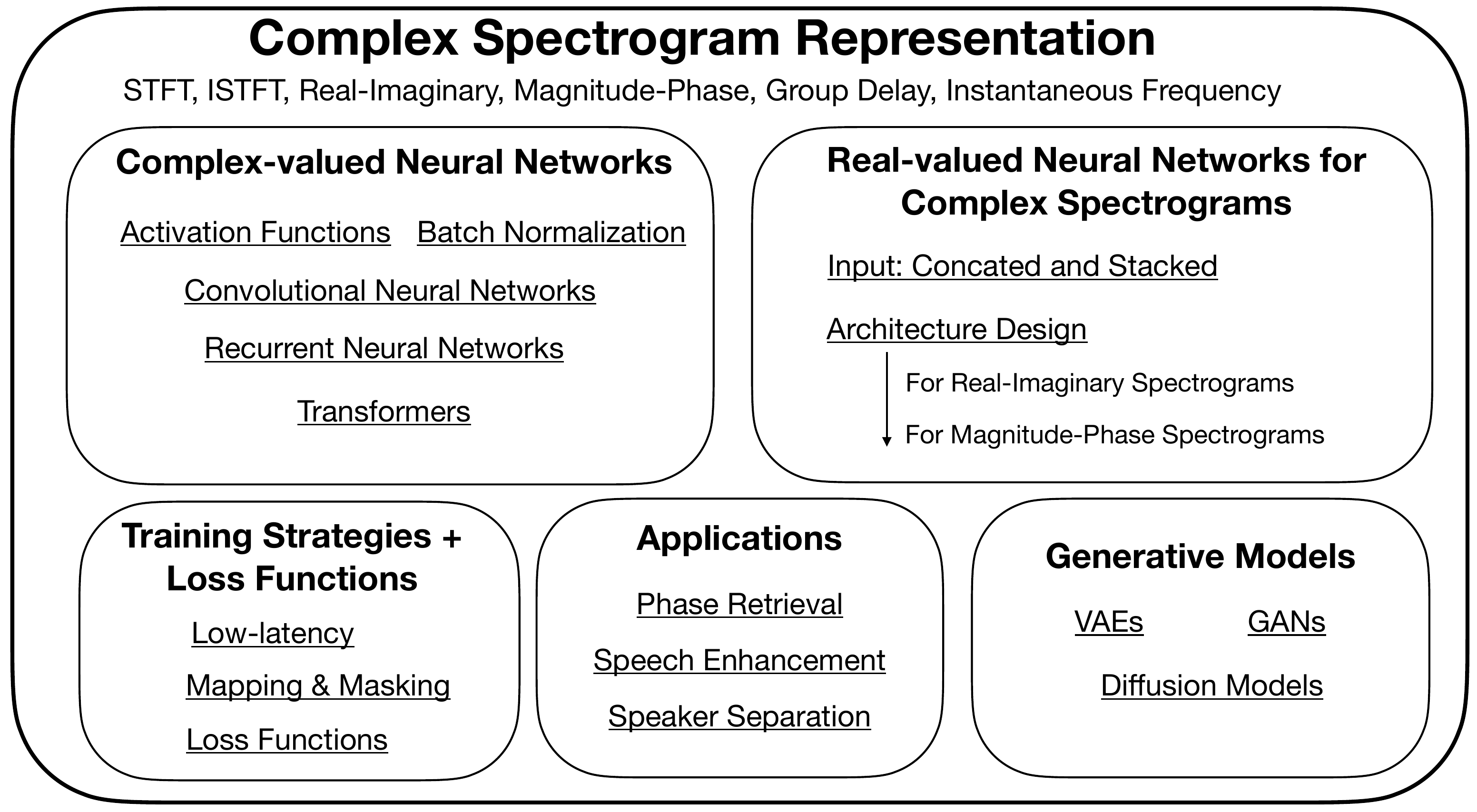}
    \caption{
    This figure illustrates the overall structure of the survey, centered on the modeling and processing of complex spectrograms. The framework is divided into five main components: (1) complex-valued neural networks specifically designed for complex spectrogram processing; (2) RVNNs applied to complex spectrograms; (3) training strategies and loss functions, encompassing low-latency methods, optimization techniques, and loss designs tailored to complex spectrogram learning; (4) applications, covering speech enhancement, speaker separation, and phase retrieval; and (5) generative models, including approaches based on VAEs, GANs, and diffusion models for complex spectrogram processing.
    }
    \label{fig:paper structure}
\end{figure*}

In recent years, deep learning has brought remarkable advancements to speech signal processing.
A substantial body of research in this field relies on short-time Fourier transform (STFT) representations, which provide detailed time-frequency analyses of speech signals. 
The STFT representations are inherently complex-valued and can be represented in either Cartesian or polar coordinate systems, resulting in real-imaginary or magnitude-phase spectrograms, respectively.

Among these representations, early studies focused predominantly on the magnitude spectrogram for several key reasons. 
First, magnitude spectrograms exhibit well-defined structural properties and facilitate effective feature extraction in deep neural network (DNN)-based approaches. 
Second, magnitude spectrograms have played a significant role in conventional speech signal processing, leading mordern DNN-based approaches to naturally inherit this focus.
In contrast, phase spectrograms have received comparatively less attention, primarily due to the challenges associated with their modeling. 
These difficulties stem from the inherent properties of phase spectrograms, including the lack of well-defined structure, the issue of phase wrapping, and a high sensitivity to temporal shifts in the waveform.
Besides, there has been skepticism regarding the significance of phase spectrograms.
Early studies, such as
\cite{phase_unimportance_ephraim_1984} and \cite{phase_unimportance_dequanwang_1982}, argued that phase spectrograms are not essential in speech processing.
However, subsequent research has led to a reassessment of this view. 
For example, \cite{phase_importance_paliwal_2011} demonstrated that enhancing phase information can improve perceptual quality.
Moreover, \cite{intro_phase_important_jonathen_2011signal} showed that combining a magnitude spectrogram with an irrelevant phase yields an unexpected audio waveform.
These findings underscore the critical role of phase information and have spurred growing interest in incorporating phase modeling into modern approaches.
Moreover, reconstruction of waveforms from estimated magnitude spectrograms commonly relies on classical algorithms such as Griffin–Lim or on neural vocoders, both of which are prone to introducing additional distortion.
This further highlights the limitations of magnitude-only approaches and emphasizes the significance of complex spectrogram processing.

Meanwhile, time-domain signals, preserving all speech information, have been widely adopted in various studies. 
The main advantages of processing time-domain signal include bypassing the STFT, thereby avoiding explicit phase spectrogram modeling and enabling low-latency processing. 
Although representative time-domain models like Conv-TasNet (\cite{LuoYi2019CSIT_convtasnet}) have shown promising performance in speaker separation, recent studies (\cite{zhang24i_interspeech_compare_TFGRID_ConvTASNET,SS_TF_GridNet_2023ICASSP}) indicate that time-frequency domain models can achieve higher performance.
Furthermore, \cite{WangZhong-Qiu2023SNSE_low_latency} observed that although Conv-TasNet employs an encoder to generate “pseudo-STFT” features, these features are not narrowband, further limiting Conv-TasNet compatibility with conventional STFT-based enhancement methods (e.g., beamforming, weighted prediction error) that rely on narrowband assumptions. 
In contrast, complex spectrogram–based models integrate seamlessly with such techniques, making them suitable for applications like multi-channel speech enhancement.
Moreover, while complex spectrogram–based methods are limited by the time–frequency resolution trade-off of the STFT and typically incur longer delays than time-domain approaches, recent studies have substantially reduced their latency.
The latest models achieve delays as low as 4 ms, enabling their deployment in real-time applications (\cite{WangZhong-Qiu2023SNSE_low_latency,latency_wuhaibin_sebastian_2025ultra}).

Increasingly, research has focused on DNN-based methods for complex spectrogram processing, 
implemented with real-valued (RVNNs) or complex-valued neural networks (CVNNs).
From the perspective of deep learning development, RVNNs remain the mainstream and have long been widely applied, while CVNNs has also garnered increasing attention in recent years.
Research on CVNNs can date back to \cite{intro_cmplx_NN_start}.
Since then, CVNNs have been applied in diverse areas such as audio processing, image denoising (\cite{intro_cmplx_NN_application_image_denoising, intro_cmplx_NN_application_medical_image_denoising}), radar signal processing (\cite{intro_cmplx_NN_application_radar1, intro_cmplx_NN_application_radar2}), and object discovery (\cite{intro_cmplx_NN_application_objective_discovery}). 
Several studies have highlighted theoretical advantages of CVNNs (\cite{intro_cmplx_NN_open_advantage1, intro_cmplx_NN_open_advantage2, intro_cmplx_NN_degree_of_freedom,intro_cmplx_NN_universal_approximation,intro_cmplx_NN_critical_points,intro_cmplx_NN_local_minima}).
Nevertheless, despite established theoretical benefits, the practical superiority of CVNNs over RVNNs remains inconclusive.
Recent studies in speech signal processing are mostly focusing on using RVNNs to handle complex spectrogram processing.
This may be because RVNNs have lower computational cost while achieving comparable performance (\cite{rethinking}).

Using deep learning for complex spectrogram processing has yielded promising results across various related fields closely linked to perceptual quality, including phase retrieval, speech enhancement, and speaker separation. 
Progress has been made in architectural design, training strategies, and loss function development throughout the exploration. 
Moreover, recent studies have increasingly focused on low-latency processing, universal models, further performance improvements, and lower computational cost, among others.
There is also growing interest in leveraging generative models for complex spectrogram processing.

However, despite these advances, the field remains fragmented and a systematic survey is still lacking. This survey aims to fill the gap and provide a comprehensive summary of DNN-based approaches for complex spectrogram processing, serving as a reference for both future research and practical applications.

This work begins by introducing the complex spectrograms and their associated features.
It then provides a summary of methods for processing complex spectrograms using CVNNs, primarily from the following perspectives:
(1) Research on CVNNs is ongoing, and our understanding of their properties and capabilities continues to evolve,
and (2) complex-valued operations possess inherent characteristics, such as the ability to represent data in both Cartesian and polar coordinates, and to simultaneously perform scaling and rotation. 
Subsequently, this review examines the use of RVNNs for complex spectrogram processing, including input handling and architectural designs. 
Rather than advocating for a particular network type, we here emphasizes that all conclusions should be grounded in experimental results and practical deployment considerations.
This work also investigates the application of deep learning to problems associated with complex spectrograms, encompassing low-latency techniques, optimization strategies, and specialized loss functions.
Additionally, we examine three key application areas — phase retrieval, speech enhancement, and speaker separation — within the framework of complex spectrogram processing. Emerging methods leveraging generative models are also summarized. 
Overall, this work aims to offer a comprehensive foundation for researchers developing DNN-based methods for complex speech spectrograms.

The structure of this survey is shown in Figure~\ref{fig:paper structure}.
In addition, Table~\ref{tab: mathematical notation summary} presents the mathematical notations used throughout this paper, while Table~\ref{tab: abbre} outlines the common abbreviations related to speech signals and deep learning referenced herein.

\begin{table}[htbp]
    \centering
    \begin{tabularx}{\linewidth}{c|p{6.5cm}}
        \hline
        \textbf{Notation} & \makecell{\textbf{Description}} \\ 
        \hline
        $\mathbb{R}$, $\mathbb{C}$ & Sets of real, and complex numbers \\
        \hline
        $x$, $\mathbf{x}$, $\mathbf{X}$ & Scalar, vector, matrix \\
        \hline
        $j$ & Imaginary unit \\
        \hline
        $\Re(x)$, $x_{\text{R}}$ & Real component of a complex-valued number $x$ \\
        \hline
        $\Im(x)$, $x_{\text{I}}$ & Imaginary component of a complex-valued number $x$ \\
        \hline
        $|x|$, $\angle x$ & Magnitude and phase of a complex-valued number $x$ \\
        \hline
        $\theta$ & Periodic variant (including phase, and phase derivatives) \\
        \hline
        $l$, $k$ & Time and frequency index in spectrogram \\
        \hline
        $\mathbb{E}$ & Expected value \\
        \hline
        $\hat{x}$ & The estimated result of variable $x$\\
        \hline
    \end{tabularx} 
    \caption{A summary of the mathematical notations used in this survey.}
    \label{tab: mathematical notation summary}
\end{table}

\begin{table}[htbp]
    \centering
    \begin{tabularx}{\linewidth}{c|p{6.5cm}}
        \hline
        \textbf{Abbreviation} & \makecell{\textbf{Description}} \\ 
        \hline
        STFT & Short-time Fourier transform \\
        \hline
        ISTFT & Inverse short-time Fourier transform \\
        \hline
        CVNN & Complex-valued neural network \\
        \hline
        RVNN & Real-valued neural network \\
        \hline
        DFT & Discrete Fourier transform \\
        \hline
        DNN & Deep neural network \\
        \hline
        IF & Instantaneous frequency \\
        \hline
        GD & Group delay \\
        \hline
        CNN & Convolutional neural network \\
        \hline
        RNN & Recurrent neural network \\
        \hline
        LSTM & Long short-term memory \\
        \hline
        MSE & Mean square error \\
        \hline
        MAE & Mean absolute error \\
        \hline
        SDR & Signal-to-distortion ratio\\
        \hline
        SSNR & Segmental signal-to-noise ratio\\
        \hline
        SI-SDR & Scale-invariant signal-to-distortion ratio\\
        \hline
        PESQ & Perceptual evaluation of speech quality\\
        \hline
        STOI & Short-time objective intelligibility\\
        \hline
        ESTOI & Extended short-time objective intelligibility\\
        \hline
        LSD & Log-spectral distortion\\
        \hline
        GLA & Griffin-Lim algorithm\\
        \hline
        GAN & Generative adversarial network\\
        \hline
        VAE & Variational autoencoder\\
        \hline
        
    \end{tabularx} 
    \caption{Common abbreviations related to speech signal processing and deep learning which are referenced throughout this paper.}
    \label{tab: abbre}
\end{table}

\section{Complex Spectrogram Representation}
\label{sec: 2. complex spectrogram representation}

\begin{figure*}[htbp]
    \centering
    \begin{subfigure}{0.33\textwidth}
        \centering
        \includegraphics[width=\linewidth]{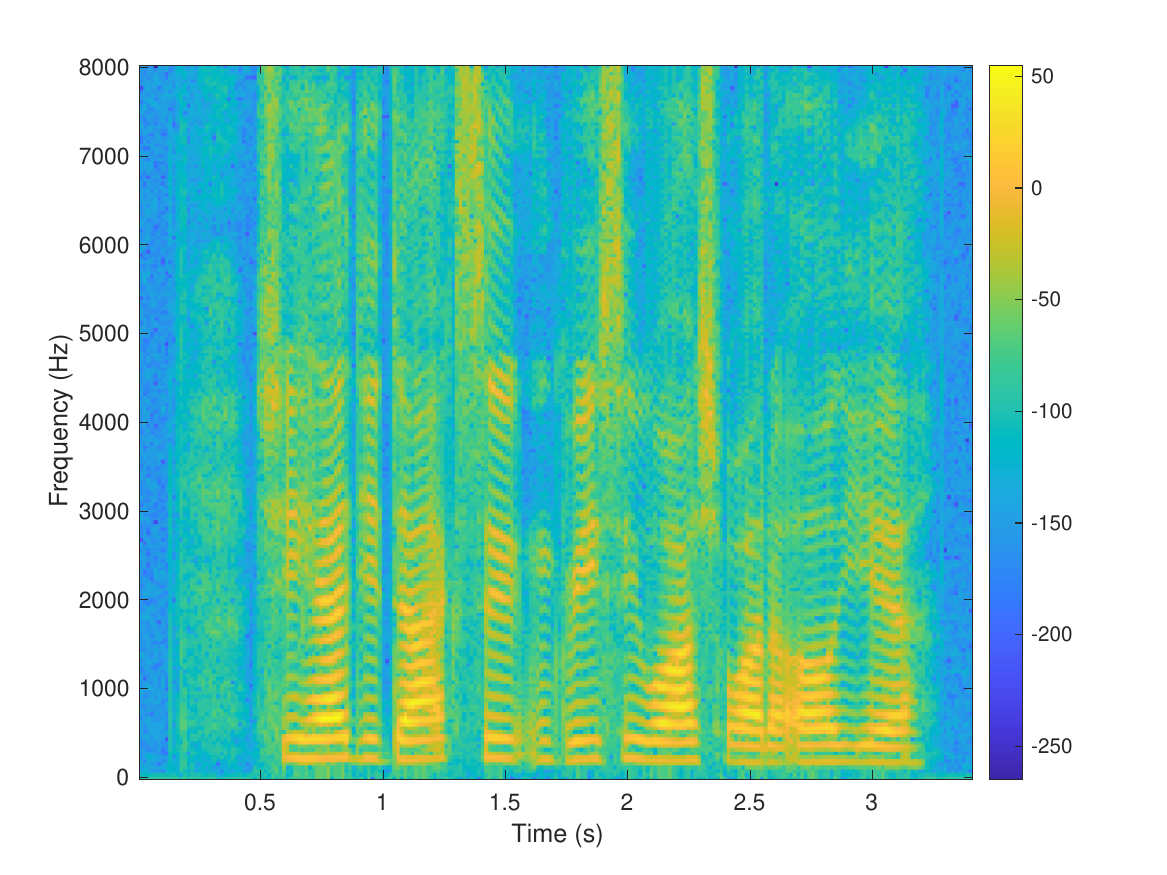}
        \caption{Log-magnitude}
    \end{subfigure}%
    \begin{subfigure}{0.33\textwidth}
        \centering
        \includegraphics[width=\linewidth]{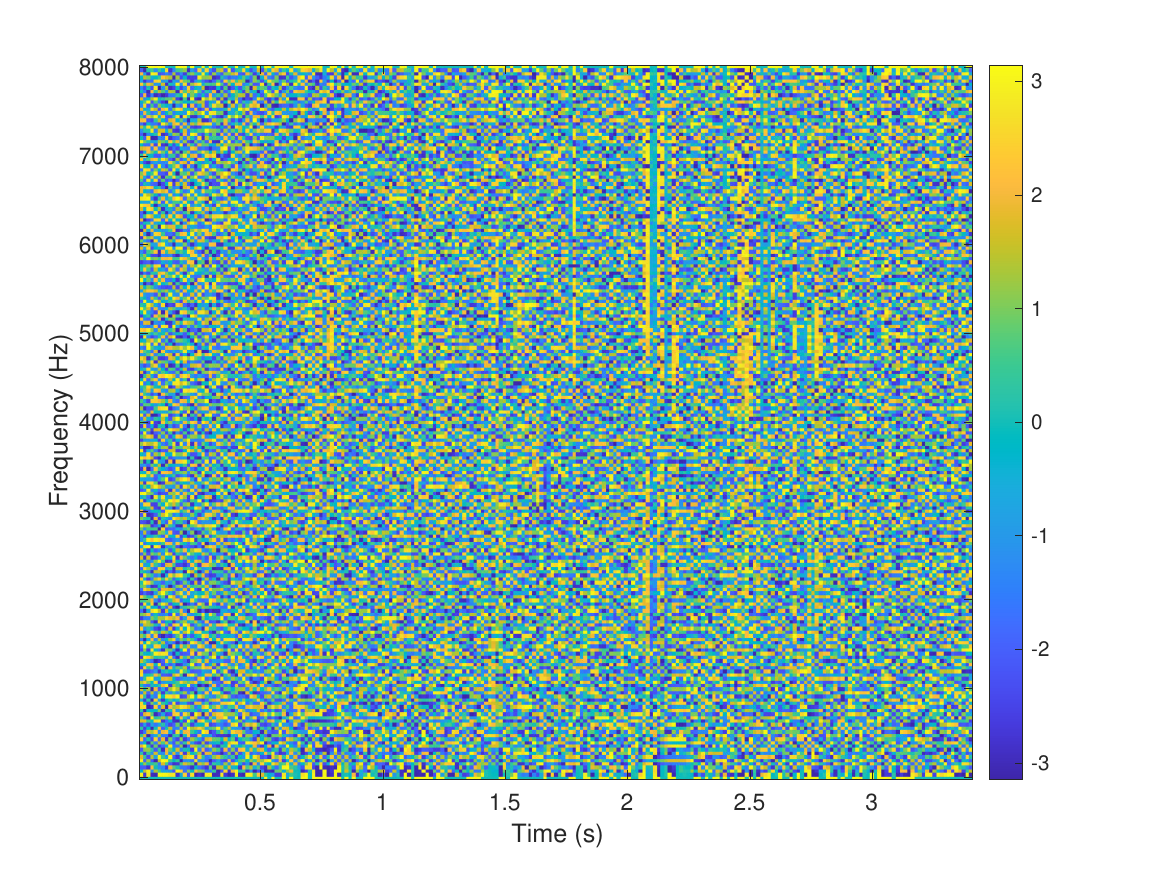}
        \caption{Phase}
    \end{subfigure}%
    \begin{subfigure}{0.33\textwidth}
        \centering
        \includegraphics[width=\linewidth]{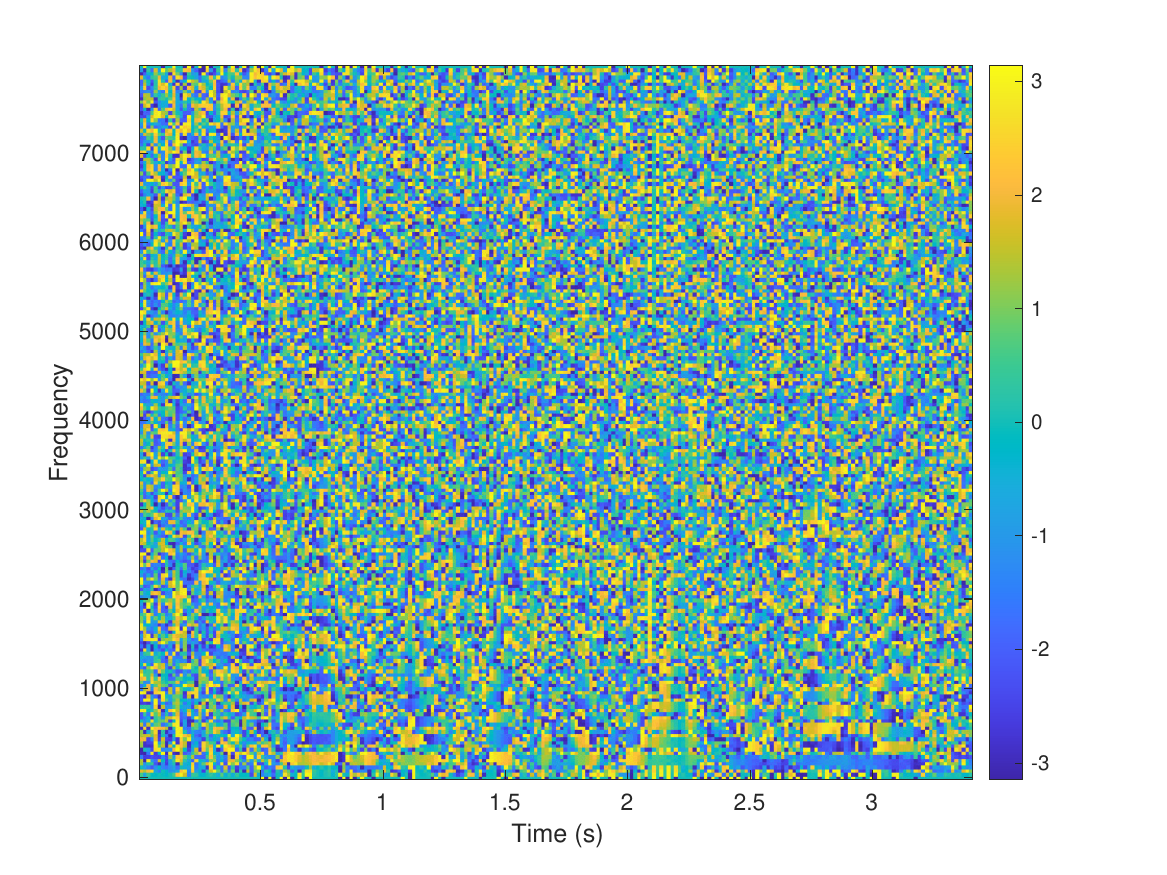}
        \caption{Group Delay}
    \end{subfigure} \\

    \begin{subfigure}{0.33\textwidth}
        \centering
        \includegraphics[width=\linewidth]{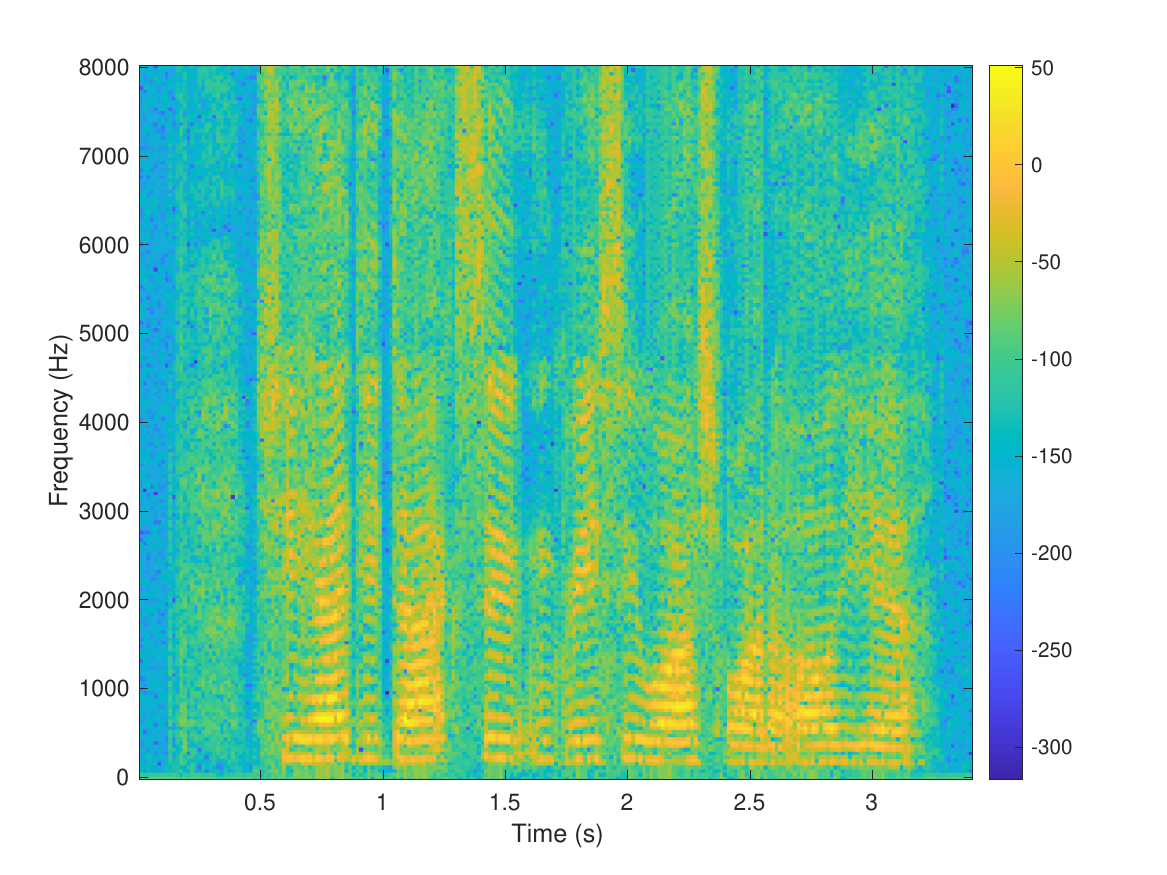}
        \caption{Real (absolute value on logarithm scale)}
    \end{subfigure}%
    \begin{subfigure}{0.33\textwidth}
        \centering
        \includegraphics[width=\linewidth]{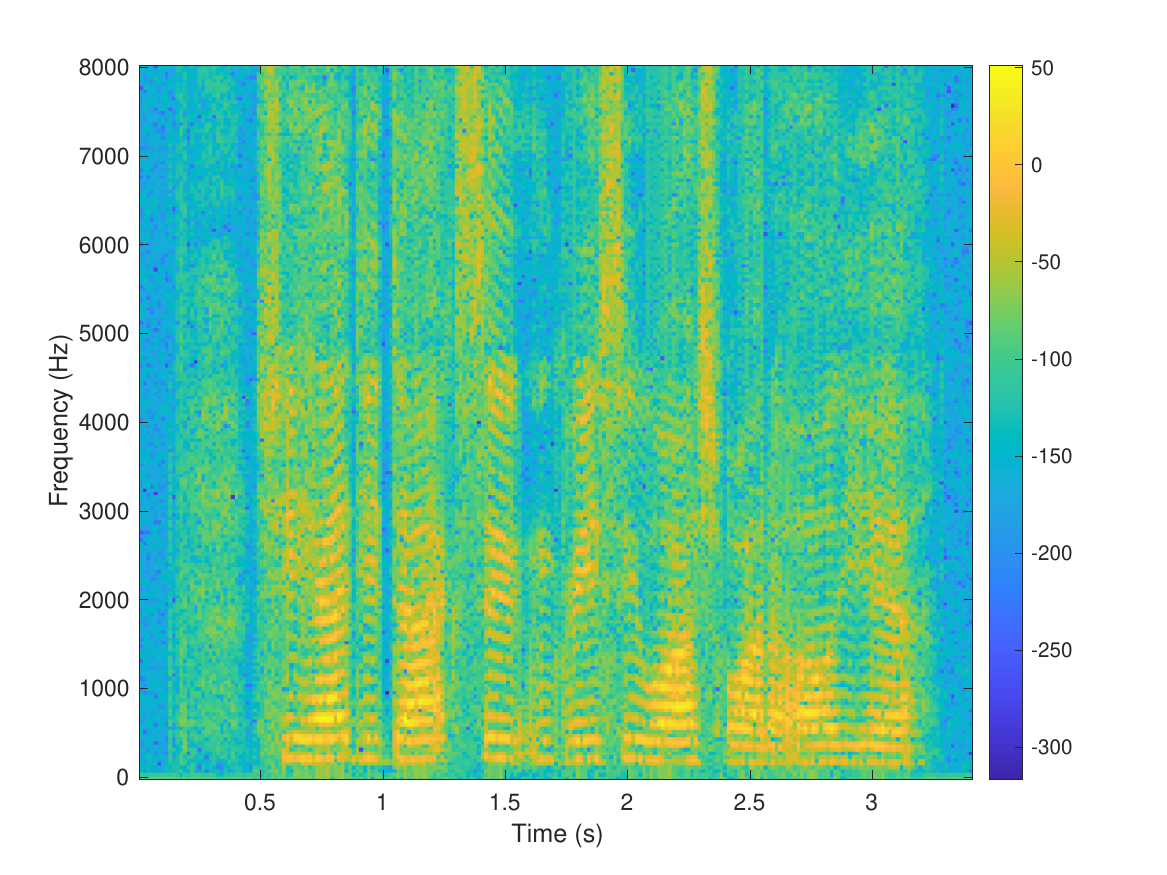}
        \caption{Imaginary (absolute value on logarithm scale)}
    \end{subfigure}%
    \begin{subfigure}{0.33\textwidth}
        \centering
        \includegraphics[width=\linewidth]{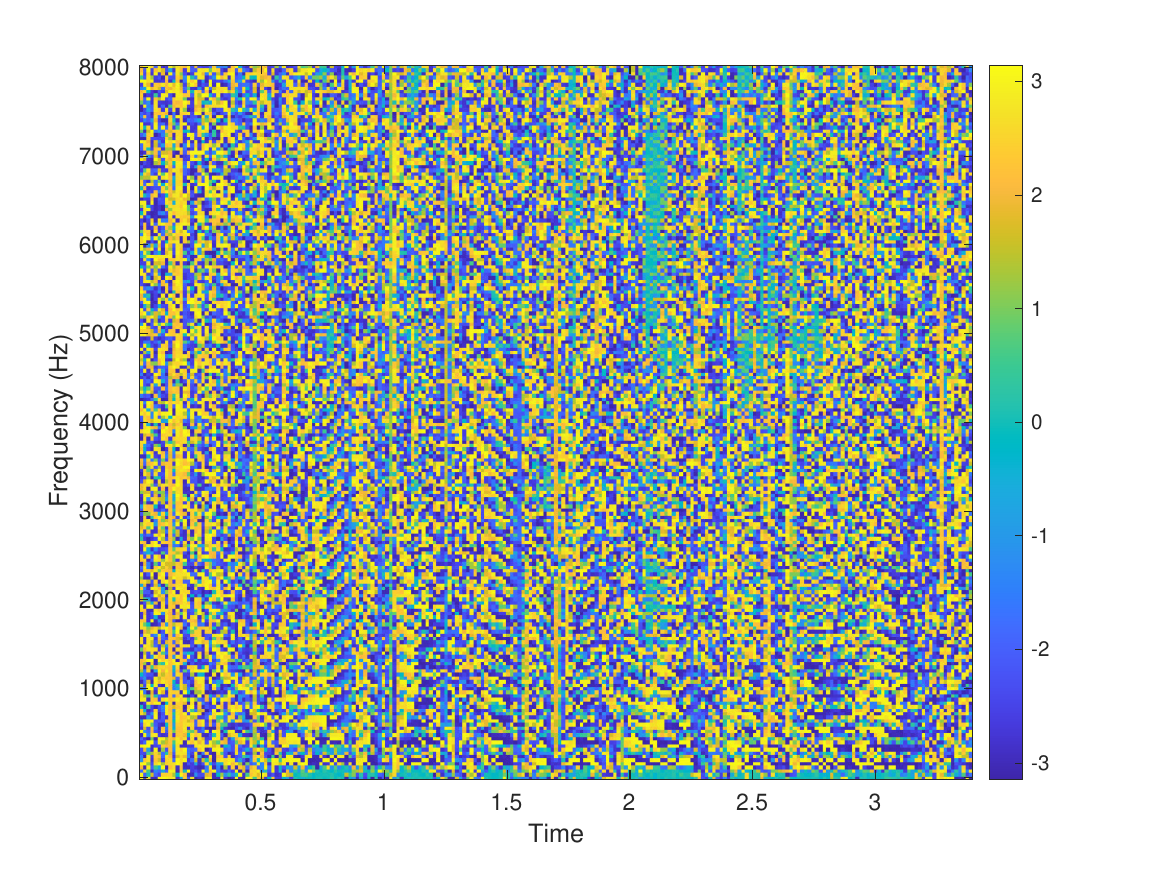}
        \caption{Instantaneous Frequency}
    \end{subfigure}

    \caption{
    Features derived from one utterance in the TIMIT dataset (speaker ID: FVMH0, utterance ID: SA1). Time-frequency representations include log-magnitude, phase, real/imaginary components, group delay, and instantaneous frequency spectrograms. Computed using STFT with a 25 ms Hamming window and 10 ms frame shift. For clarity, real and imaginary spectrograms are displayed as absolute values on the logarithmic scale.
    }
    \label{fig:spectrogram}
\end{figure*}

Assume a time sequence $\{y[n]\} \in \mathbb{R}$ as the observed signal from a monaural microphone, where $n$ is a discrete-time index. 
Its short-time Fourier transform (STFT) representation can be calculated as:
\begin{equation}
Y(l,k) = \sum_{m=0}^{M-1}y(lH+m)w(m)\exp(-j2\pi mk /M).
\end{equation}
where $M$ indicates the number of points in a frame taken from $y(n)$ (i.e. frame length),  $H$ the frame shift length, 
$w(m)$ the window used in the Fourier transform, $l$ the time index and $k$ the frequency axis.

The STFT representation $Y(l,k)$ is naturally complex-valued, and can be illustrated 
in Cartesian coordinate system with real and imaginary components:
\begin{equation}
    Y(l,k)=\Re(Y(l,k)) + j \Im(Y(l,k)),
\end{equation}
or in the polar coordinate system with magnitude and phase:
\begin{equation}
    Y(l,k)=|Y(l,k)|\exp(j\theta_{Y}(l,k)),
\end{equation}
where $j$ is the imaginary unit.
Hence, the complex spectrogram itself offers distinct representations for subsequent deep learning algorithms.


However, the phase component poses unique challenges for deep learning.
As shown in Figure~\ref{fig:spectrogram}, which presents the log-magnitude, phase, real, and imaginary spectrograms from a TIMIT utterance, clear structures are visible in the log-magnitude, real, and imaginary components, while the phase spectrogram appears more chaotic.
Due to their structured patterns, real and imaginary spectrograms have become widely used in deep learning, enabling more effective modeling by DNNs (\cite{cIRM_TASLP}).
They are now commonly applied across tasks such as enhancement, separation, and generative modeling.

Research working on magnitude-phase spectrograms is also ongoing, but challenges are mostly posed in phase spectrogram processing.
This is typically attributed to the fact that phase is wrapped into the range of $[-\pi, \pi]$, and it is hard to find clear structure.
Therefore, derivatives of phase are often used to bypass directly dealing with phase itself.
The derivatives of phase spectrogram $\theta(l,k)$ over time is called instantaneous frequency (IF; \cite{IF_stark2008speech}):
\begin{equation}
    \theta_{\text{IF}} = \theta(l,k) - \theta(l-1,k),
\end{equation}
and over frequency, which is referred to as group delay (GD; \cite{GD_yegnanarayana1992significance}),
\begin{equation}
    \theta_{\text{GD}} = \theta(l,k) - \theta(l,k-1).
\end{equation}
Compared to the phase spectrogram, both IF and GD exhibit more distinct and interpretable structural patterns, as illustrated in Figure~\ref{fig:spectrogram}. 
This characteristic has led to their adoption as input features and as features used in the formulation of regularization terms.
A key research focus in processing magnitude-phase spectrograms is phase retrieval, which aims to reconstruct the phase from its corresponding magnitude.
Moreover, the magnitude-phase representation has been studied in a wide range of speech processing applications, particularly in speech enhancement and separation tasks.


Given the known complex spectrogram $\tilde{X}(l,k)$, the inverse short time Fourier transform (ISTFT) is apllied to obtain the time-domain estimated signal $\tilde{x}[n]$. 
The first step of ISTFT is an inverse discrete Fourier transform (DFT) on each frame:
\begin{equation}
    \begin{aligned}
        \tilde{x}_l(m) = \frac{1}{M} \sum_{k=0}^{M-1} \tilde{X}(l,k) \exp{(j2\pi mk/M)}.
    \end{aligned}
\end{equation}
Then the overlap-add operation is used to get the time-domain signal $\tilde{x}[n]$ via:
\begin{equation}
    \begin{aligned}
        \tilde{x}[n] = \sum_{l} v[n-lH]\tilde{x}_l[n-lH],
    \end{aligned}
\end{equation}
in which $v[n]$ is the synthesis window.

\section{Key Components of Complex-Valued Neural Networks}
\label{sec: 3. Key Components of Neural Networks}

In this section, we will briefly introduce the complex-valued activation functions and complex-valued batch normalization that appear in CVNNs when dealing with complex spectrograms.


\subsection{Activation Functions}
Assume that $z=\Re(z)+j\Im(z)$ is a complex-valued variable. Below are several complex-valued activation functions applied to $z$.

\subsubsection{ModReLU}
ModReLU, a variation of ReLU adapted for the complex domain, was first proposed by \cite{uRNN}. This point-wise nonlinear function is defined as follows:
\begin{equation}
\label{eq: modReLU - 3.1 Activation Functiosn}
    \sigma_{modReLU}(z) = \sigma_{ReLU}(|z|+b)\frac{z}{|z|} = \left \{ 
    \begin{array}{lr}
    (|z|+b)\frac{z}{|z|}  & \text{if} \, |z| + b \geq 0 \\
    0  &\text{if} \, |z| + b < 0
    \end{array}
    \right .
\end{equation}
The input $z \in \mathbb{C}$, while the bias parameter $b \in \mathbb{R}$.
Additionally, $\sigma_{ReLU}(\cdot)$ denotes the real-valued ReLU function.

\subsubsection{$\mathbb{C}$\textnormal{ReLU}}
As first introduced by~\cite{deep_complex_networks}, the complex ReLU ($\mathbb{C}$\textnormal{ReLU}) applies the real-valued ReLU activation function to both real and imaginary parts independently:
\begin{equation}
\label{eq: CReLU - 3.1 Activation Functiosn}
    \mathbb{C}\textnormal{ReLU}(z) =\textnormal{ReLU}(\Re(z)) + j\textnormal{ReLU}(\Im(z)).
\end{equation}
$\mathbb{C}$\textnormal{ReLU} satisfies the Cauchy-Riemann functions when $\theta_z\in (0,\pi/2)$ or $\theta_z \in (\pi, 3\pi/2)$, i.e. both the real and imaginary components are strictly positive or negative.
A variant of $\mathbb{C}$ReLU, leaky $\mathbb{C}$ReLU has been used in complex U-Net (\cite{complex_unet}) by replacing the ReLU in Eq.~\eqref{eq: CReLU - 3.1 Activation Functiosn} with leaky ReLU for more stable training.

\subsubsection{$z$\textnormal{ReLU}}
Another ReLU-based complex activation function was proposed by~\cite{ZReLU} and is referred to as $z$ReLU in~\cite{deep_complex_networks}.
The definition of $z$ReLU is as below: 
\begin{equation}
\label{eq: ZReLU - 3.1 Activation Functiosn}
    z\textnormal{ReLU} =\left \{
    \begin{aligned}
        & z \quad & \text{if} \, \theta_z \in [0, \pi/2] \\
        & 0 \quad & \text{otherwise}
    \end{aligned}
    \right.
\end{equation}
This activation function satisfies the Cauchy-Riemann equations everywhere except for the set of points $\{\Re(z)>0, \Im(z)=0\}\cup\{\Re(z)=0, \Im(z)>0\}$.

\subsubsection{ModSigmoid}
\label{Actiation: modSigmoid}
The modSigmoid function, proposed by~\cite{cGRU}, is defined as follows:
\begin{equation}
\label{eq: ModSigmoid - 3.1 Activation Functiosn}
    f_{modSigmoid}(z) = \sigma(\alpha \Re(z)+\beta \Im(z)), \quad \alpha,\beta \in [0,1], 
\end{equation}
where $\sigma(\cdot)$ denotes the real-valued sigmoid function, and  $\alpha$ and $\beta$ are trainable variants.

\subsection{Complex-Valued Batch Normalization}

Complex batch normalization was first proposed by~\cite{deep_complex_networks}.
Let $\mathbf{x}$ denote a  2D complex-valued vector. 
The formulation of complex batch normalization can be described as follows:
\begin{equation}
\label{eq: BN definition - 3.2 Batch Normalization}
    \text{BN}(\tilde{\mathbf{x}}) = \bm{\gamma} \Tilde{\mathbf{x}} + \bm{\beta},
\end{equation}
where
\begin{equation}
\label{eq: mean - 3.2 Batch Normalization}
    \tilde{\mathbf{x}} = (\mathbf{V})^{-\frac{1}{2}}(\mathbf{x}-\mathbb{E}[\mathbf{x}]),
\end{equation}
and $\mathbb{E}[\mathbf{x}]$ represents the mean of the vector $\mathbf{x}$.
The covariance matrix $\mathbf{V}$ is defined as:
\begin{equation}
\label{eq: covariance matrix - 3.2 Batch Normalization}
    \mathbf{V}= \left(
    \begin{array}{cc}
        V_{\text{RR}} & V_{\text{RI}} \\
        V_{\text{IR}} & V_{\text{II}}
    \end{array} 
    \right)
    = \left(
    \begin{array}{cc}
    \text{Cov}(\Re(\mathbf{x}), \Re(\mathbf{x})) &  \text{Cov}(\Re(\mathbf{x}), \Im(\mathbf{x}))\\
    \text{Cov}(\Im(\mathbf{x}), \Re(\mathbf{x})) & \text{Cov}(\Im(\mathbf{x}), \Im(\mathbf{x}))
    \end{array}
    \right).
\end{equation}
The trainable scaling parameter $\bm{\gamma}$ is defined as:
\begin{equation}
\label{eq: scaling parameter - 3.2 Batch Normalization}
    \bm{\gamma} = \left( 
    \begin{array}{cc}
        \gamma_{\text{RR}} & \gamma_{\text{RI}} \\
        \gamma_{\text{RI}} & \gamma_{\text{II}} 
    \end{array}
    \right).
\end{equation}
In the initial work of \cite{deep_complex_networks}, the diagonal elements $\gamma_{\text{RR}}$ and $\gamma_{\text{II}}$ are initialized to $\frac{1}{\sqrt{2}}$, and $\gamma_{\text{RI}}$ to 0.
Furthermore, the elements of the shift parameter $\bm{\beta}$ are initialized to 0+0j.

\section{Complex-Valued Neural Networks}
\label{sec: 4. complex-valued neural networks}

In this section, we will introduce CVNNs, covering various layer types, including feed-forward neural networks, recurrent neural networks, convolutional neural networks, and transformers.
For details on how these layers are integrated into complete architectures (e.g., DCUNET, DCCRN), please refer to Section 7.

\subsection{Complex-Valued Feed-forward Neural Networks}
For a complex-valued feed-forward neural network, the weights and bias are denoted as $\mathbf{W}\in \mathbb{C}^{N\times M}$ and $\mathbf{b} \in \mathbb{C}^{N \times 1}$.
A single layer of the complex-valued feed-forward neural network can then be expressed as:
\begin{equation}
    \begin{aligned}
        \mathbf{h} = f(\mathbf{W}\mathbf{x} + \mathbf{b}),
    \end{aligned}
\end{equation}
where $\mathbf{x} \in \mathbb{C}^{M\times 1}$ is the input and $\mathbf{h} \in \mathbb{C}^{N\times 1}$ is the output of the network. 
Additionally, $f(\cdot)$ represents a complex-valued activation function.

\subsection{Complex-Valued Convolutional Neural Networks}
\label{sec: 4.2 complex-valued convolutional neural networks}

\cite{deep_complex_networks} introduced a complex-valued convolutional layer.
Assuming both the filter weight matrix $\mathbf{W}=\Re(\mathbf{W})+j\Im(\mathbf{W})$ and the input vector $\mathbf{x}=\Re(\mathbf{x})+j\Im(\mathbf{x})$ are complex-valued, the complex-valued convolution is expressed as:
\begin{equation}
\label{eq 4.2: complex convolutional}
\begin{aligned}
    \mathbf{W}*\mathbf{x}&
    =(\Re(\mathbf{W})*\Re(\mathbf{x})-\Im(\mathbf{W})*\Im(\mathbf{x}))\\
    &+j(\Im(\mathbf{W})*\Re(\mathbf{x})+\Re(\mathbf{W})*\Im(\mathbf{x})).
\end{aligned}
\end{equation}
Meanwhile, drawing inspiration from the success of real-valued weight initialization methods (\cite{xavier_init, he_init}), \cite{deep_complex_networks} further proposed complex-valued initialization methods for weights.
Assuming a complex-valued weight $w$ whose magnitude $|w|$ follows a Rayleigh distribution with scale parameter $\sigma$, after derivation, the variance of $w$ can be expressed as:
\begin{equation}
    \text{Var}(w)=2\sigma^2.
\end{equation}
To align with Xavier Initialization (\cite{xavier_init}), the variance of the complex weight is set as $\text{Var}(w)=2/(n_{in}+n_{out})$ , leading to $\sigma=1/\sqrt{n_{in}+n_{out}}$, where 
$n_{in}$ and $n_{out}$ represent the number of input and output units, respectively.
Similarly, to follow Kaiming Initialization (\cite{he_init}), the variance is defined as $\text{Var}(w)=2/n_{in}$, resulting in $\sigma=1/\sqrt{n_{in}}$.
The initialization of the complex-valued weights is then performed by sampling the magnitude from a Rayleigh distribution with the scale parameter $\sigma$ and the phase from a uniform distribution $\mathcal{U}[-\pi, \pi]$.


Subsequently, \cite{complex_unet} proposed an alternative implementation of Eq.~\eqref{eq 4.2: complex convolutional} by leveraging real-valued convolutional neural network (CNN) operations.
In this approach, the real and imaginary components of the weight matrix, $\Re(\mathbf{W})$ and $\Im(\mathbf{W})$, are implemented by two separate real-valued CNN layers.
Despite this decomposition, the calculation mechanism still adheres to the complex-valued convolution rule outlined in Eq.~\eqref{eq 4.2: complex convolutional}.
Real-valued initialization techniques, such as Xavier or normal initialization, are commonly used to separately initialize $\Re(\mathbf{W})$ and $\Im(\mathbf{W})$ in this manner (\cite{xavier_init, complex_unet, hu2020dccrn}).

\subsection{Complex-Valued Recurrent Neural Networks}


Complex-valued recurrent neural networks (RNNs) were initially proposed to address notorious vanishing and exploding gradient problems, which are particularly severe in long sequence processing.
One solution to these issues in real-valued RNNs is to use orthogonal weight matrices, leveraging its norm-preserving property (\cite{CVRNN_orthogonal_weights1, CVRNN_orthogonal_weights2}).
Specifically, if a matrix $\mathbf{W}$ is orthogonal, it satisfies $\|\mathbf{W}h\|_2 = \|h\|_2$, as $\mathbf{W}^{\intercal}\mathbf{W}=\mathbf{W}\mathbf{W}^{\intercal}=\mathbf{I}$, where $\intercal$ denotes matrix transpose.
This norm-preserving property is particularly beneficial for RNNs, as it helps to preserve gradients over time, mitigating the vanishing or exploding gradient problem.

Extending this idea to the complex domain, unitary matrices can be employed to achieve similar gradient-preserving benefits.
Building on this principle, \cite{uRNN} proposed the unitary evolution RNN (uRNN), which constrains the recurrent weight matrix to be unitary.
The norm-preserving property is retained in uRNN, while the complexity of parameterizing orthogonal matrices is alleviated by leveraging the complex domain.
Experimental results have demonstrated the effectiveness of uRNN on long-term dependency tasks.
Subsequently, several studies have addressed the limitations and proposed various extensions of uRNN (\cite{FC-uRNN, ceRNN}).

Given that previous studies on complex-valued RNNs have shown positive results, \cite{cGRU} were the first to specifically investigate a complex RNN architecture that operates directly on complex-valued spectrograms.
\cite{cGRU} first defined a basic formulation for a complex RNN as:
\begin{equation}
    \mathbf{z}_t=\mathbf{W}\mathbf{h}_{t-1}+\mathbf{V}\mathbf{x}_t+\mathbf{b},
\end{equation}
\begin{equation}
    \mathbf{h}_t=f_a(\mathbf{z}_t),
\end{equation}
where $\mathbf{x}_t\in \mathbb{C}^{n_x\times 1}$ represents the input vector at time step $t$, and $\mathbf{h}_t\in \mathbb{C}^{n_h\times 1}$ denotes the hidden unit vector.
The dimensions of the input and hidden states are denoted by $n_x$ and $n_h$, respectively.
The input-to-state and hidden-to-hidden transitions are parameterized by the weight matrices $\mathbf{V}\in \mathbb{C}^{n_h\times n_x}$ and $\mathbf{W}\in \mathbb{C}^{n_h\times n_h}$, respectively, while $\mathbf{b}\in \mathbb{C}^{n_h \times 1}$ represents the bias term.
They further proposed the complex gated RNN (cgRNN) as:
\begin{equation}
    \tilde{\mathbf{z}}_t = \mathbf{W}(\mathbf{g}_r\odot\mathbf{h}_{t-1})+\mathbf{V}\mathbf{x}_t + \mathbf{b},
\end{equation}
\begin{equation}
    \mathbf{h}_t=\mathbf{g}_z\odot f_a(\tilde{\mathbf{z}}_t) + (1-\mathbf{g}_z)\odot \mathbf{h}_{t-1},
\end{equation}
where $\mathbf{g}\odot\mathbf{h}=\mathbf{g}\odot|\mathbf{h}|e^{j\theta_h}$ is the element-wise product.
Reset and update gates are denoted as $\mathbf{g}_r$ and $\mathbf{g}_z$ with definitions: 
\begin{equation}
    \mathbf{g}_r = f_g(\mathbf{z}_r), \quad \mathbf{z}_r=\mathbf{W}_r\mathbf{h}+\mathbf{V}_r\mathbf{x}_t + \mathbf{b}_r,
\end{equation}
\begin{equation}
    \mathbf{g}_z = f_g(\mathbf{z}_z), \quad \mathbf{z}_z=\mathbf{W}_z\mathbf{h}+\mathbf{V}_z\mathbf{x}_t + \mathbf{b}_z.
\end{equation}
The gate activation function $f_g(\cdot)$ is the modSigmoid given in Eq.~\eqref{eq: ModSigmoid - 3.1 Activation Functiosn}.
The state-to-state transition matrices are denoted as $\mathbf{W}_r,\mathbf{W}_z\in \mathbb{C}^{n_h\times n_h}$.
The input-to-state transition matrices are denoted as $\mathbf{V}_r, \mathbf{V}_z\in \mathbb{C}^{n_h \times n_i}$. 
The biases are $\mathbf{b}_r, \mathbf{b}_z\in \mathbb{C}^{n_h}$.



Another line of research in complex-valued recurrent neural networks is inspired by real-valued RNN architectures.
Building on the real-valued long short-term memory (LSTM), a quasi-complex-valued LSTM was proposed by (\cite{hu2020dccrn}) and has been widely adopted.

Assuming the complex input is given by $\mathbf{X}=\Re(\mathbf{X}) + j \Im(\mathbf{X})$, the quasi-complex-valued LSTM layer is computed as follows:
\begin{equation}
\begin{aligned}
    &\mathbf{F}_{\text{RR}} = \text{LSTM}_{\text{R}}(\Re(\mathbf{X})), \quad &\mathbf{F}_{\text{IR}} = \text{LSTM}_{\text{R}}(\Im(\mathbf{X})), \\
    &\mathbf{F}_{\text{RI}} = \text{LSTM}_{\text{I}}(\Re(\mathbf{X})), \quad &\mathbf{F}_{\text{II}} = \text{LSTM}_{\text{I}}(\Im(\mathbf{X})),
    \end{aligned}
\end{equation}
where $\text{LSTM}_{\text{R}}$ and $\text{LSTM}_{\text{I}}$ represent two real-valued LSTM layers, respectively.
The real and imaginary components of the input are processed separately by these two layers.
The final output $\mathbf{F}_{\text{out}}$ of the quasi-complex LSTM is obtained as:
\begin{equation}
\label{eq: quasi LSTM out}
    \mathbf{F}_{\text{out}} = (\mathbf{F}_{\text{RR}}-\mathbf{F}_{\text{II}}) + j(\mathbf{F}_{\text{RI}}+\mathbf{F}_{\text{IR}}).
\end{equation}

Additionally, an LSTM layer similar to the traditional real-valued LSTM but with fully complex-valued calculations was used for comparison with the quasi-complex-valued LSTM in (\cite{rethinking}).
However, as this fully complex-valued LSTM did not yield satisfactory performance, we do not elaborate on its details here.

\subsection{Complex-Valued Transformers}
As the Transformer model (\cite{real_trans}) has demonstrated its powerful capability in processing sequential data (\cite{transformer_importance_ASR}), several works have emerged to extend the attention layer to the complex-valued domain.

\cite{complex_trans_muqiao_yang} first proposed a complex-valued transformer, by leveraging the real-valued attention layers.
Assume the complex-valued input as $\mathbf{X}=\Re(\mathbf{X})+j\Im(\mathbf{X})=\mathbf{X}_{\text{R}} + j \mathbf{X}_{\text{I}}$.
The definition of a complex multi-head attention is then as:
\begin{equation}
\begin{aligned}
\label{eq: complex attention layer - multihead X}
    &\text{ComplexMultiAttention}(\mathbf{X}) \\
    = &(\text{MultiHead}(\mathbf{X}_{\text{R}},\mathbf{X}_{\text{R}},\mathbf{X}_{\text{R}})-\text{MultiHead}(\mathbf{X}_{\text{R}},\mathbf{X}_{\text{I}},\mathbf{X}_{\text{I}}) \\
    &-\text{MultiHead}(\mathbf{X}_{\text{I}},\mathbf{X}_{\text{R}},\mathbf{X}_{\text{I}}) - \text{MultiHead}(\mathbf{X}_{\text{I}},\mathbf{X}_{\text{I}},\mathbf{X}_{\text{R}})) \\
    &+ j(\text{MultiHead}(\mathbf{X}_{\text{R}},\mathbf{X}_{\text{R}},\mathbf{X}_{\text{I}})+ \text{MultiHead}(\mathbf{X}_{\text{R}},\mathbf{X}_{\text{I}},\mathbf{X}_{\text{R}})\\
    &+\text{MultiHead}(\mathbf{X}_{\text{I}},\mathbf{X}_{\text{R}},\mathbf{X}_{\text{R}})-
    \text{MultiHead}(\mathbf{X}_{\text{I}},\mathbf{X}_{\text{I}},\mathbf{X}_{\text{I}})),
\end{aligned}
\end{equation}
where $\text{MultiHead}(\cdot)$ denotes a real-valued multi-head attention mechanism (\cite{real_trans}). 
Min-Max-Normalization is incorporated as the activation function in $\text{MultiHead}(\cdot)$ to enhance training stability, expressed as:
\begin{equation}
\begin{aligned}
\label{eq: complex attention layer - multihead QKV}
    &\text{MultiHead}(\mathbf{Q},\mathbf{K},\mathbf{V})\\
    =& \text{Concat}(\{\text{Attention}(\mathbf{QW}_i^Q, \mathbf{KW}_i^K, \mathbf{VW}_i^V)\}_{i=1}^n)\mathbf{W},
\end{aligned}
\end{equation}
\begin{equation}
    \text{Attention}(\mathbf{Q},\mathbf{K},\mathbf{V})=\text{Min-Max-Norm}(\frac{\mathbf{QK}^T}{\sqrt{d_k}})\mathbf{V},
\end{equation}
\begin{equation}
    \text{Min-Max-Norm}(\mathbf{Z})=\frac{\mathbf{Z}-\min(\mathbf{Z})}{\max(\mathbf{Z})-\min(\mathbf{Z})}.
\end{equation}
Factor $d_k$ is the dimension of $\mathbf{Q}$ and $\mathbf{K}$.
\cite{complex_trans_muqiao_yang} utilized Xavier uniform as the initializer of the complex transformer.

In addition, several studies have proposed various attention mechanisms inspired by or based on the real-valued attention mechanism (\cite{complex_trans_tongji,complex_trans_muster, complex_trans_jesper}).
For instance, \cite{complex_trans_jesper} proposed a complex transformer block for binaural speech enhancement.
The output $\mathbf{H}^{n+1}=\mathbf{H}^{n+1}_{\text{R}}+j\mathbf{H}^{n+1}_{\text{I}}$, is calculated separately for the real and imaginary components:
\begin{equation}
    \mathbf{H}^{n+1}_{\text{R}} = \text{MultiHead}(\mathbf{H}^{n}_{\text{R}}, \mathbf{H}^{n}_{\text{R}}) - \text{MultiHead}(\mathbf{H}^{n}_{\text{I}}, \mathbf{H}^{n}_{\text{I}})
\end{equation}
\begin{equation}
    \mathbf{H}^{n+1}_{\text{I}} = \text{MultiHead}(\mathbf{H}^{n}_{\text{R}}, \mathbf{H}^{n}_{\text{I}}) + \text{MultiHead}(\mathbf{H}^{n}_{\text{I}}, \mathbf{H}^{n}_{\text{R}})
\end{equation}
where $\text{MultiHead}(\cdot)$ is the real-valued multi-head attention operator from~\cite{real_trans}.


\section{Using Real-Valued Neural Networks for Complex Speech Spectrograms}
\label{sec: add5. real-valued neural networks}

Compared with CVNNs, RVNNs are more mature and widely studied.
In this section, we briefly introduce approaches that employ RVNNs for complex spectrogram processing.
Since a complex spectrogram can be represented in either the Cartesian coordinate system (real and imaginary components) or the polar coordinate system (magnitude and phase components), different strategies have been proposed.
We begin by examining the inputs to real-valued neural networks, then proceed to the architectural designs for handling the real and imaginary components of spectrograms, and finally discuss architectures that operate on magnitude and phase representations.

\subsection{Input Representations: Concatenated and Stacked}
\begin{figure}[htbp]
    \centering
    \begin{subfigure}{0.35\columnwidth}
        \centering
        \includegraphics[width=\columnwidth]{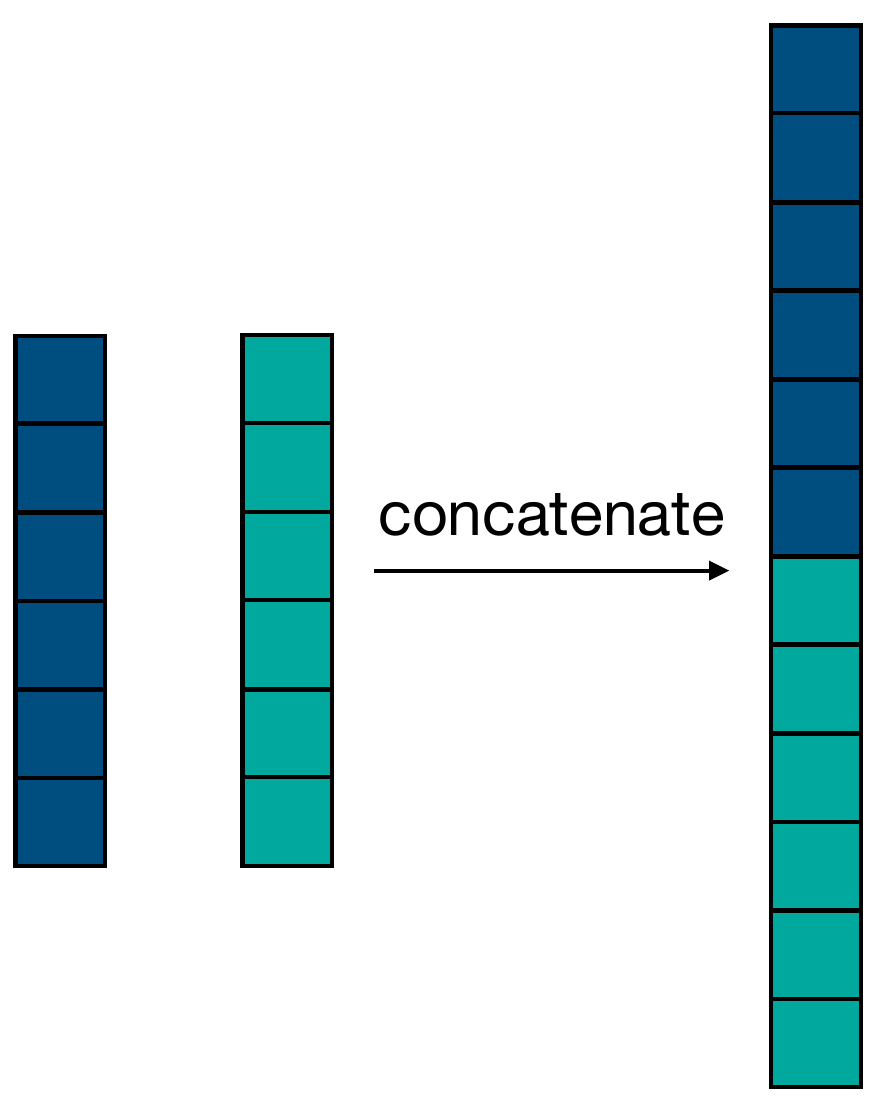}
        \caption{Concatenated}
    \end{subfigure}%
    \hspace{5em}
    \begin{subfigure}{0.35\columnwidth}
        \centering
        \includegraphics[width=\columnwidth]{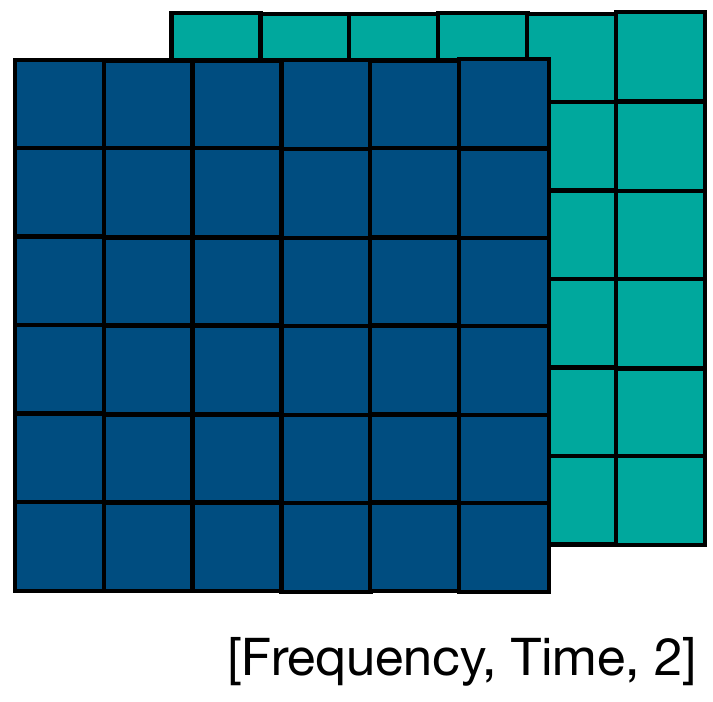}
        \caption{Stacked}
    \end{subfigure}%
    \caption{
    Illustration of two common input representations for real-imaginary or magnitude-phase spectrograms in RVNN architectures.
    (a) Concatenated representation (Concatenated): The real and imaginary (or magnitude and phase) components of each complex-valued time–frequency vector are concatenated into a single vector, resulting in a higher-dimensional real-valued feature vector.
    (b) Stacked representation (Stacked): the real and imaginary (or magnitude and phase) components are stacked into a three-dimensional tensor with dimensions [Frequency, Time, 2], where the last dimension corresponds to the real and imaginary (or magnitude and phase) components.
    }
    \label{fig:RVNN-input}
\end{figure}

Depending on the architectural design, the front-end of RVNNs typically process the complex spectrograms using one of the two main strategies.

The first strategy is the concatenated vector representation, where the real-imaginary or magnitude-phase components of each complex-valued time–frequency vector are concatenated into a single higher-dimensional vector before being fed into the network, as illustrated in Fig.~\ref{fig:RVNN-input}(a).
Representative works adopting this approach include \cite{SE_CIRM_williamson_2017taslp}, \cite{Cmplx_sound_localization2}, FullSubNet (\cite{hao2021fullsubnet}), USR-BSRNN (\cite{RI_concat_USR-BSRNN}), and MambAttention (\cite{RI_concat_mambattention}).

The second strategy is the stacked representation, as shown in Fig.~\ref{fig:RVNN-input}(b). In this case, a complex spectrogram of size $[Frequency, Time]$ is extended to $[Frequency, Time, 2]$, where the additional dimension corresponds to the real and imaginary components or the magnitude and phase spectrograms. 
The stacked tensor is then treated as a multi-channel real-valued input, analogous to the RGB channels of an image. 
Representative works include \cite{fu_szu_wei_multi_task_complex_MLSP_2017}, CRN (\cite{SE_CRN_ke_tan_2019taslp}), CMGAN (\cite{SE_generative_CMGAN}) and MP-SENet (\cite{Training_strategy_masking+mapping_MP_SENet}).

Overall, the concatenated vector representation is commonly used in architectures with a fully connected layer or RNN layer as the front end (i.e., the first layer is fully connected or RNN), whereas the stacked representation arranges the spectrogram components channel-wise — treating the real, imaginary or magnitude, phase parts as separate components — and is therefore particularly suitable for architectures with a convolutional front end. 
Subsequent layers (e.g., CNN, RNN or attention layers) can be incorporated into the framework to further improve performance.

\subsection{Architectural Design}
\subsubsection{For Real-Imaginary Spectrograms}
In early efforts to process real-imaginary spectrograms using RVNNs, fully connected layers were directly applied (\cite{SE_CIRM_williamson_2017taslp}). \cite{fu_szu_wei_multi_task_complex_MLSP_2017} were among the first to introduce a CNN framework for processing complex speech signals.

Speech signals exhibit clear temporal characteristics, with strong dependencies between contextual frames, and uneven energy distribution across frequency bands. 
Therefore, subsequent research has often adopted methods that combine global and local information to enhance model performance. 
For example, CRN (\cite{SE_CRN_ke_tan_2019taslp}) employs a U-Net-like encoder-decoder structure with skip connections across the encoder and decoder, and incorporates RNN-based blocks to capture global temporal dependencies. 
Similar improved models include PoCoNet (\cite{RVNN_SE_poconet}) and DPCRN (\cite{SE_DPCRN_Xiaohuai_Le_2021interspeech}).

Moreover, modeling of full-band and sub-band information has also received widespread attention. 
Full-band models can capture global information but lack dedicated mechanisms for modeling sub-band features, such as signal stationarity and local spectral patterns. 
Conversely, subband models (\cite{RVNN_SE_subband}) are the exact opposite.
FullSubNet (\cite{hao2021fullsubnet}) addresses this issue by sequentially connecting full-band and sub-band models and jointly training them to integrate the advantages of both. 

Following the dual-path RNN architecture (\cite{SE_orig_dual_path_RNN}), many subsequent models have leveraged both sub-band and cross-band modeling to more effectively capture temporal and spectral information.
In contrast to FullSubNet which sequentially connects full-band and sub-band models for joint learning, dual-path approaches typically integrate a sub-band module and a cross-band module within a single block. 
This block is then stacked multiple times to progressively refine the representations.
Representative models include DPT-FSNet (\cite{SE_DPT-FSNet_ICASSP2022}), CMGAN (\cite{SE_generative_CMGAN}), TF-GridNet (\cite{SS_TF_GridNet_2023ICASSP}), SpatialNet (\cite{SS_quan2024spatialnet}), and TF-CrossNet (\cite{SS_kalkhorani2024tf_TF_CrossNet}), which effectively capture fine-grained local spectral details while simultaneously modeling global time–frequency dependencies.

Besides, the inherent spectral energy imbalance across frequencies can lead to distortion and information loss in high-frequency components. 
To address this, the Band-Split RNN (BSRNN) was proposed (\cite{BSRNN_orig}). 
In this approach, the input signal is split into several sub-bands. A bi-directional LSTM is employed to process information in the lower-frequency bands ($<$8 kHz), while a uni-directional LSTM is used to capture high-frequency information ($>$8 kHz).

More recently,increasing attention has been directed toward the development of universal models that can robustly process audio signals under diverse conditions, such as varying sampling rates and multi-channel configurations. 
Notable examples include USR-BSRNN (\cite{RI_concat_USR-BSRNN}) and \cite{unviersal_wangyou_zhang_asru2023}, among others.

\subsubsection{For Magnitude-Phase Spectrograms}
\label{subsec: RVNN for magnitude-phase}

\begin{figure}[htbp]
    \centering
\begin{subfigure}{0.8\columnwidth}
        \centering
        \includegraphics[width=\columnwidth]{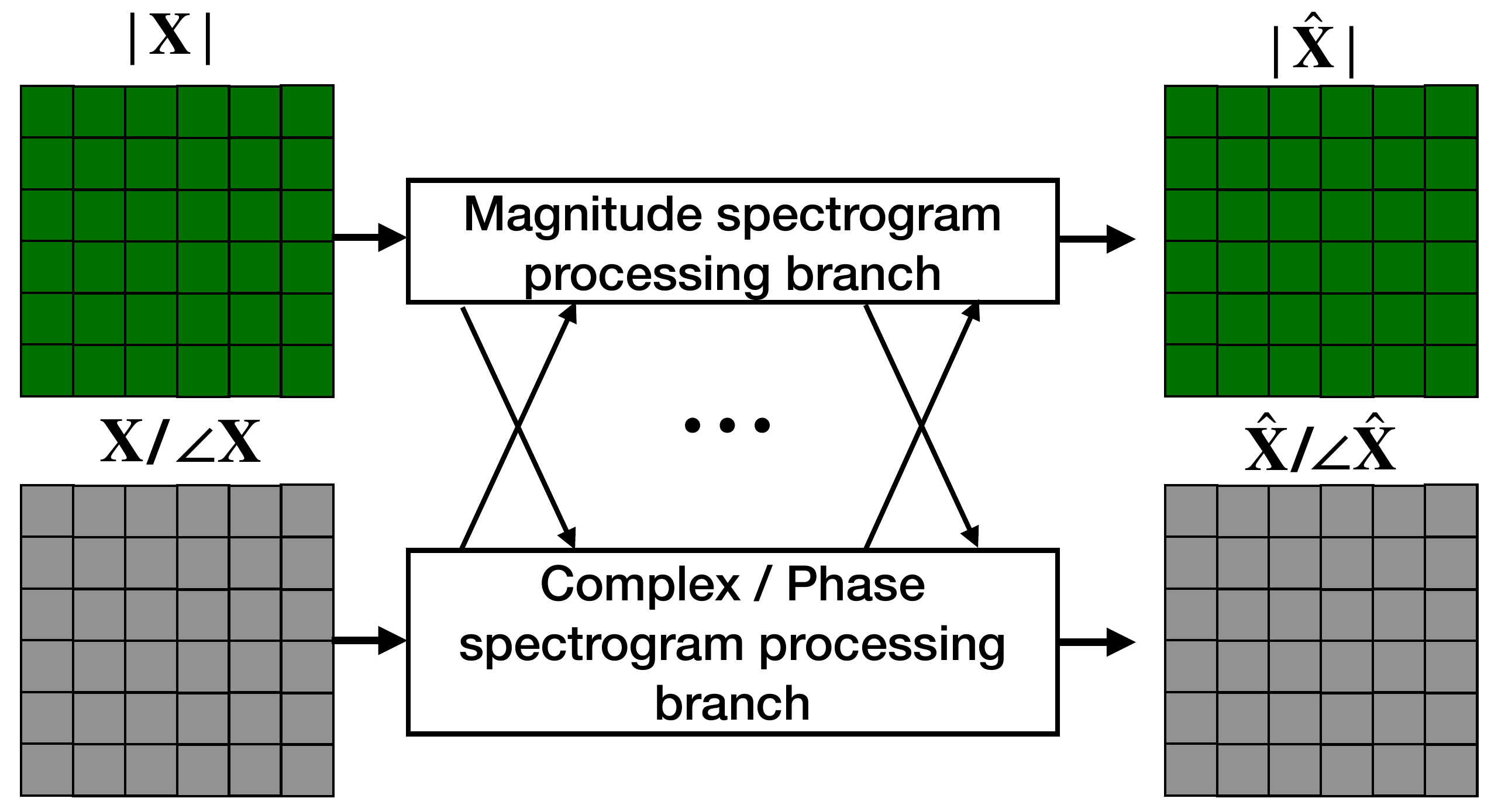}
        \caption{Dual-Branch}
    \end{subfigure}%
    \hspace{4em}
    \begin{subfigure}{\columnwidth}
        \centering
        \includegraphics[width=\columnwidth]{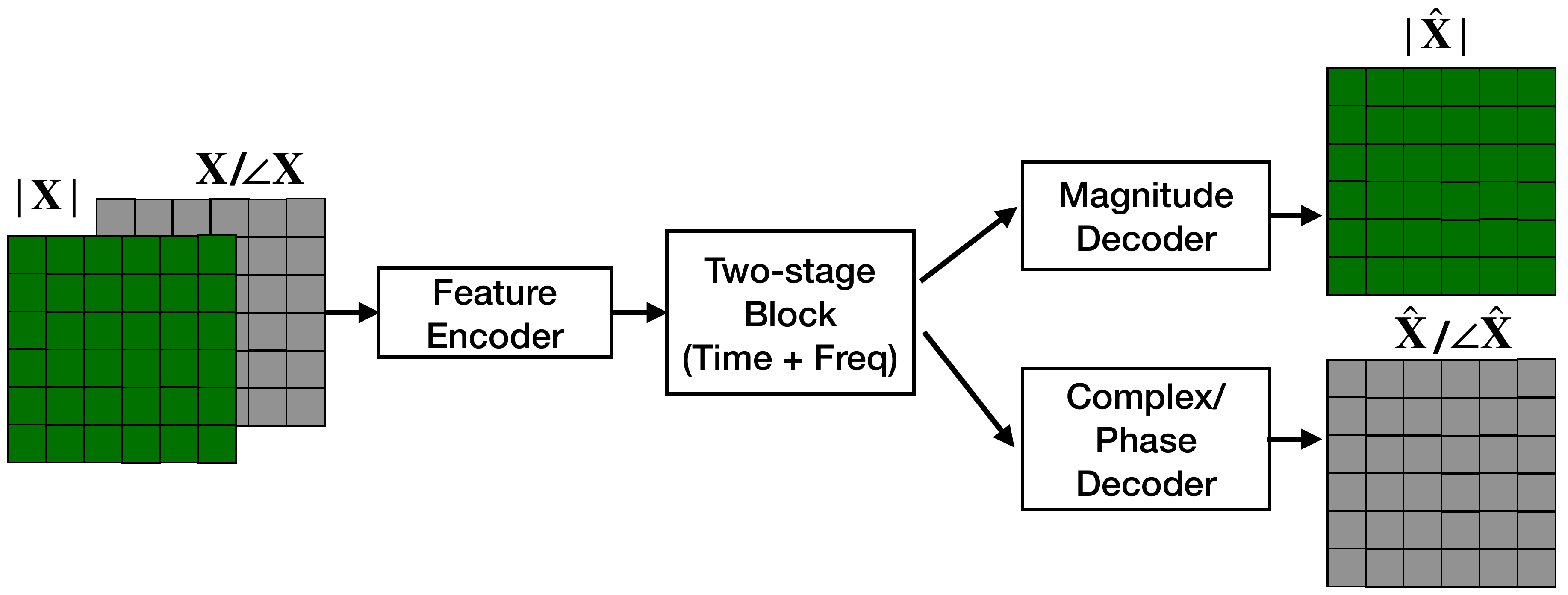}
        \caption{Encoder-Decoder}
    \end{subfigure}%
    \caption{
    Illustration of two commonly used architectures for processing magnitude and phase spectrograms.
    }
    \label{fig:RVNN-input-mag-pha}
\end{figure}

RVNNs for processing magnitude and phase spectrograms often adopt a dual-branch (also referred to as dual-path\footnote{Two distinct "dual-path" mechanisms have been proposed in the literature. The first is exemplified by the dual-path RNN, which incorporates intra- and inter-chunk processing to capture both local and global information. 
Representive works include dual-path RNN (\cite{SE_orig_dual_path_RNN}) and TF-GridNet (\cite{SS_TF_GridNet_2023ICASSP}) among others.  
The second dual-path mechanism, noted in this work as dual-branch, discussed here, employs a parallel architecture, where magnitude and phase information are processed in separate branches. 
}) architecture, as shown in Fig~\ref{fig:RVNN-input-mag-pha} (a). 
In this framework, one branch is dedicated to modeling the magnitude spectrogram, while the other focuses on the complex (\cite{SE_Uformer_ICASSP2022}) or phase (\cite{PHASEN}) spectrogram. 
Since phase information is inherently more difficult to model, effective interaction between the two branches is typically required, and multiple information exchanges are often introduced across different layers to facilitate joint learning.
Representative works under this line include PHASEN (\cite{PHASEN}), UFormer (\cite{SE_Uformer_ICASSP2022}), 
and DB-AIAT(\cite{SE_DB_AIAT_ICASSP2022}).

Recently, there exists an encoder-decoder-based architecture for processing magnitude-phase spectrograms, as shown in Fig~\ref{fig:RVNN-input-mag-pha} (b).
Within this framework, both magnitude and phase spectrograms are first encoded into feature representations by an encoder. 
These features are then fed into a two-stage block, which processes them sequentially along the temporal and frequency dimensions. 
The block is typically instantiated with models such as Conformer (\cite{Conformer}), Mamba (\cite{Mamba}), or xLSTM (\cite{xlstm}). 
Finally, the processed features are passed to the magnitude decoder and phase decoder, where the corresponding spectrogram components are reconstructed.
Representative works include MPSENet (\cite{Training_strategy_masking+mapping_MP_SENet}) and MambaAttantion (\cite{RI_concat_mambattention}).

\section{Training Strategies and Loss Functions}
\label{sec: 5. training strategies and loss functions}

For real-time speech applications such as hearing aids and telecommunication systems, low latency is essential. 
Conventional spectrogram-based methods often require long window length in STFT calculation for better performance, but at the cost of additional algorithm latency.
To address this, approaches such as asymmetric windowing and future-frame prediction have been explored.
In addition, when DNNs are used to solve problems related to complex spectrograms, the main goal is to estimate the target complex spectrograms from the input.
The estimation approaches vary, and a range of different training strategies have been employed.
These strategies can generally be categorized into direct estimation (mapping) and indirect estimation (masking).
Finally, loss functions are employed to measure how closely the algorithm’s output matches the target.
Since complex spectrograms can be represented in various coordinate systems, multiple loss functions have been proposed and studied.
The first subsection will summarize the low-latency techniques that enable practical deployment.
The second subsection below will provide an overview of the training strategies employed in complex spectrogram-related works.
The third subsection will discuss the loss functions utilized for learning complex spectrogram-related targets.

\subsection{Low-latency}
\label{sec: sub: low-latency}

Low-latency is a critical concern, especially in real-time speech processing systems where preserving perceptual quality is essential. 
Here latency refers to the time delay between the input of a speech signal and the production of the processed output.
Latency generally consists of two components: algorithmic latency and hardware-induced computational latency (\cite{latency_google_dementyev2025towards,WangZhong-Qiu2023SNSE_low_latency, latency_FBE_zheng2022low}).

The STFT window length largely determines the algorithmic latency in complex spectrogram–based speech processing. 
Owing to the inherent trade-off between time and frequency resolution, the choice of window length directly affects model performance. 
Shorter windows reduce latency but yield poor frequency resolution, thereby limiting the capability of DNN models.
Conversely, longer windows provide higher frequency resolution and richer spectral details, which further enhance the modeling capacity of DNN frameworks, but often fail to satisfy the strict latency requirements of real-time applications (\cite{latency_wuhaibin_sebastian_2025ultra, WangZhong-Qiu2023SNSE_low_latency}).

To enable real-time processing with high speech quality, several strategies have been developed for handling complex spectrograms (\cite{latency_wuhaibin_sebastian_2025ultra}).
One approach is to use asymmetric STFT windowing (\cite{MaulerDirk2007Aldv_low_latency,wang2021deepneuralnetworkbased_low_latency,WangZhong-Qiu2023SNSE_low_latency}).
Here, a larger analysis window is used to compute the input STFT features, while a shorter synthesis window is employed for the ISTFT. 
This approach has been shown to reduce latency in complex spectrogram-based models to as low as 4 ms (\cite{WangZhong-Qiu2023SNSE_low_latency}).

Latency can also be reduced through future-frame prediction (\cite{latency_mads, WangZhong-Qiu2023SNSE_low_latency}). 
This technique predicts the features of the next frame directly from past and current observations, eliminating the need to wait for the actual future frame. 
With a 50\% overlap in STFT calculation, the total latency can be effectively reduced to the hop size. 
Moreover, for a given latency constraint, future-frame prediction allows the use of a longer analysis window, thereby improving the frequency resolution of the input spectrograms.

\subsection{Mapping and Masking}

The primary objective of deep learning-based complex spectrogram-related works is to accurately estimate the target complex spectrograms through neural network optimization. 
Various strategies have been employed to achieve more precise estimation results.
These strategies are generally classified into direct and indirect estimation.

The direct estimation paradigm, which implements DNNs to establish non-linear input-target mappings through parametric function approximation, is formally designated as mapping-based approaches in relevant literatures.
Mapping-based approaches have been employed across multiple complex spectrogram manipulation tasks, including phase retrieval, speech enhancement and separation.

Indirect estimation, which refers to estimating the target complex spectrogram in an indirect manner, is a widely adopted paradigm in speech processing.
In speech-related tasks, this approach is typically realized through masking-based methods.
These methods fundamentally conceptualize deep neural networks (DNNs) as adaptive spectral filters.
Specifically, a neural network is trained to estimate a time-frequency mask, which is then applied to the input complex spectrogram via element-wise multiplication to extract the target components.
Masking-based techniques are predominantly employed in tasks such as speech enhancement and separation.
In the following, we review several representative masks that are either phase-sensitive or directly applicable to complex spectrogram processing.

The first one is phase-sensitive masking (PSM), which was proposed by \cite{erdogan2015PSM}.
The definition of PSM is:
\begin{equation}
 M_{\text{PSM}} = \frac{|X|}{|Y|} \cos(\angle \theta_X - \angle \theta_Y),
\end{equation}
where $X$ and $Y$ represent the clean and input speech spectrogram bins, and $\theta_X$ and $\theta_Y$ denote their respective phases.
To avoid non-negative values in PSM, \cite{mortern_PIT} proposed Ideal Non-negative PSM (INPSM) defined as:
\begin{equation}
    M_{\text{INPSM}} = \max (0, M_{\text{PSM}}).
\end{equation}
Both $M_{\text{PSM}}$ and $M_{\text{INPSM}}$ are real-valued masks that multiply the magnitude of the noisy spectrogram, while the noisy phase is retained during the synthesis stage.

The complex ideal ratio mask (cIRM) (\cite{cIRM_TASLP, cIRM_ICASSP}), as an extension of the ideal ratio mask, is complex-valued.
Assume the input signal spectrogram bin $Y=Y_{\text{R}}+jY_{\text{I}}$, and the target signal spectrogram bin as $X=X_{\text{R}}+jX_{\text{I}}$, the cIRM is then calculated as:
\begin{equation}
    M_{\text{cIRM}}=\frac{Y_{\text{R}} X_{\text{R}} + Y_{\text{I}} X_{\text{I}}}{Y_{\text{R}}^2 + Y_{\text{R}}^2} + j \frac{Y_{\text{R}} X_{\text{I}} - Y_{\text{I}} X_{\text{R}}}{Y_{\text{R}}^2 + Y_{\text{I}} ^2}.
\end{equation}

The multiplication between a complex-valued masking $M$ and input spectrogram $Y$ can be considered in either the rectangular coordinate system (\cite{hao2021fullsubnet}) as:
\begin{equation}
    M \times Y = (\Re(M) + j \Im(M))(\Re(Y) + j \Im(Y)), 
\end{equation}
or the polar coordinate system (\cite{complex_unet,hu2020dccrn}) as:
\begin{equation}
    M \times  Y = |M| |Y| \exp{(\theta_{M} + \theta_Y)}.
\end{equation}
Although complex multiplication is mathematically identical across coordinate systems, practical implementations - such as the use of activation functions - often vary (\cite{hu2020dccrn}), leading to subtle differences between the two approaches.

The masking-based methods have demonstrated their advantages in previous real-valued works (\cite{wang2014training_masking, michelsanti2021overview}).
This could be because, compared to mapping, generating a mask is easier for neural networks, as filtering noisy speech is less challenging than synthesizing clean speech (\cite{gabbay18_interspeech, michelsanti2019loss_comparison}).
Motivated by these advantages, many studies in complex spectrogram processing have also adopted masking-based approaches for tasks such as speech enhancement and separation (\cite{cIRM_TASLP, complex_unet, hu2020dccrn}).
However, recent findings suggest that mapping-based methods can also achieve strong performance in complex domains (\cite{rethinking}).
In this paper, we refrain from drawing an absolute conclusion in complex-valued spectrogram processing, as the comparison between mapping and masking approaches may depend on the network architecture and training settings.

Moreover, recent research has witnessed the emergence of integrated approaches combining masking and mapping techniques, particularly in the context of magnitude and phase spectrogram processing (\cite{PHASEN, Training_strategy_masking+mapping_MP_SENet, Training_strategy_masking+mapping_xLSTM_SENet}).
Building on the success in magnitude spectrogram processing, masking remains central in these frameworks — responsible for magnitude estimation, while a parallel branch handles phase mapping.

\subsection{Loss Functions}

A well-designed loss function can substantially improve model performance without increasing computational overhead, thereby serving as a critical component in the training process.
Given that complex spectrograms can be represented in multiple coordinate systems and subsequently transformed back to the time domain, a variety of loss functions have been proposed to align with these distinct representations. 
In this survey, we systematically present the loss functions from four perspectives: phase-related loss functions, complex-domain loss functions, time-domain loss functions, and multi-domain loss functions.
The discussion of magnitude-domain loss functions is omitted, as they have been extensively studied and this survey focuses on complex spectrogram-realted works.


\subsubsection{Phase-Related Loss Functions}
\label{sec: 5.2.1 phase-related loss functions}
Estimating phase spectrograms is a challenging task due to their intrinsic properties, such as the lack of distinct structure and the complexities introduced by phase wrapping.
Research related to phase has been explored in various domains, including phase retrieval (\cite{phase_retrieval_cosine_loss}), speech enhancement (\cite{PHASEN}), and speaker separation (\cite{phase_loss_speech_separation}).
Details about the natural properties of phase and recent advances in phase retrieval will be discussed in Subsection~\ref{sec: 6. phase retrieval}.
In the following, we outline the phase-related loss functions that are frequently employed.

Due to the periodic nature of phase, the cosine loss function has gained prominence in recent phase retrieval research (\cite{phase_retrieval_cosine_loss, phase_retrieval_stable_cosine_loss_two_stage}).
The study by \cite{phase_retrieval_cosine_loss} proposed the cosine loss function,  which is derived from the log-likelihood of a von Mises distribution, as:
\begin{equation}
    \mathcal{L}(\bm{\theta},\hat{\bm{\theta}}) = \sum_{l,k} - \cos(\theta_{l,k}-\hat{\theta}_{l,k}).
\end{equation}
Here, $\bm{\theta}$ represents a periodic variant, which may refer not only to the phase itself but also to its derivatives.
The indices $k$ and $l$ correspond to the frequency bin and time frame, respectively.
Furthermore, \cite{phase_retrieval_stable_cosine_loss_two_stage} proposed a regularized cosine loss
which yields more stable training. 

Additionally, to avoid phase wrapping in loss calculation, \cite{phase_retrieval_anti_wrapping_zhenhua_ling} proposed loss functions defined as:
\begin{equation}
    \mathcal{L}_{\text{IP}}=\mathbb{E}_{\bm{\theta},\hat{\bm{\theta}}}[\|f_{\text{AW}}(\theta-\hat{\theta})\|_1],
\end{equation}
\begin{equation}
    \mathcal{L}_{\text{GD}}=\mathbb{E}_{\Delta_{\text{DF}}(\bm{\theta},\hat{\bm{\theta}})}[\|f_{\text{AW}}(\Delta_{\text{DF}}(\theta-\hat{\theta}))\|_1],
\end{equation}
\begin{equation}
    \mathcal{L}_{\text{IF}}=\mathbb{E}_{\Delta_{\text{DT}}(\bm{\theta},\hat{\bm{\theta}})}[\|f_{\text{AW}}(\Delta_{\text{DT}}(\theta-\hat{\theta}))\|_1].
\end{equation}
The true and estimated phase spectrograms are denoted by $\bm{\theta}$ and $\hat{\bm{\theta}}$, respectively.
The anti-wrapping function $f_{\text{AW}}(\cdot)$ is defined as $f_{\text{AW}}(t)=|t-2\pi \cdot \text{round}(t/2\pi)|$, where $t\in \mathbb{R}$. 
$\Delta_{\text{DF}}$, $\Delta_{\text{DT}}$ represent the derivatives along the frequency and time axes, respectively.
The final phase loss function from \cite{phase_retrieval_anti_wrapping_zhenhua_ling} is defined as $\mathcal{L}_{\text{Pha}}=\mathcal{L}_{\text{IP}}+\mathcal{L}_{\text{GD}}+\mathcal{L}_{\text{IF}}$.

\subsubsection{Complex-Domain Loss Functions}

The $L^p$ norm applied to the real and imaginary components is the commonly used complex-domain loss, and can be formulated as:
\begin{equation}
\label{eq: RI loss}
    \mathcal{L}_{\text{RI}} = \| \Re(\hat{\mathbf{X}})- \Re(\mathbf{X})\|_p + \| \Im(\hat{\mathbf{X}})- \Im(\mathbf{X})\|_p,
\end{equation}
where $\mathbf{X}$ and $\hat{\mathbf{X}}$ represent the ground truth and the estimated results, respectively.
The choice of the L-p norm, which can be either L1 or L2, varies across different studies.
The mean square error (MSE) has long been used for log-magnitude spectrograms (\cite{MSE_mag_1985}), and its complex-valued form is also widely adopted for complex spectrograms (\cite{SE_CIRM_williamson_2017taslp, strake2019separated_L2_complex, nakashika2020_CVAE}).
However, \cite{braun2021consolidated} conducted a comparative study between the L1 norm and L2 norm as described in Eq. (\ref{eq: RI loss}).
Their findings revealed that the distributions of the real and imaginary components of the spectrogram align more closely with a Laplacian distribution rather than a Gaussian distribution.

Moreover, the spectrogram often exhibits an imbalanced energy distribution across different frequency bands.
Specifically, higher energy tends to be concentrated in the lower frequency range, while higher frequencies are characterized by relatively weaker energy.
Such disparity may lead to frequency-dependent biases in the neural network's output.
One effective approach is to apply compression to the magnitude spectrogram, helping to equalize energy levels across different frequency bands (\cite{Andong_Li_JASA_importance_of_compression}).
Mathematically, the compression process can be expressed as:
\begin{equation}
\label{eq: compressed spectrogram}
    \mathbf{X}_{\text{C}} = |\mathbf{X}|^{\beta} \exp{(\boldsymbol{\theta}_\mathbf{X})},
\end{equation}
where $\mathbf{X}_{\text{C}}$ represents the compressed complex spectrogram. 
In this formulation, the original magnitude spectrogram $|\mathbf{X}|$ is compressed with a power coefficient $\beta$, commonly set to 0.3 in the literature, while the phase spectrogram $\boldsymbol{\theta}_\mathbf{X}$ is preserved in the compressed output.
This compressed spectrogram then serves as the feature $\mathbf{X}$ in Eq. (\ref{eq: RI loss}) to compute the complex-domain loss.
The experimental results from \cite{Andong_Li_JASA_importance_of_compression} demonstrate the advantage of this power compression method over other approaches, such as logarithmic compression.

As the complex spectrogram can be represented in either Cartesian coordinates or polar coordinates, some works impose distinct penalties on the magnitude and phase spectrograms (\cite{phase_aware_xiaolei_zhang_TASLP_2019, PHASEN, Training_strategy_masking+mapping_MP_SENet, real_VAE_complex_spec_2019_ICASSP}).
These loss functions can be collectively represented as:
\begin{equation}
\label{eq: magnitude-phase loss}
    \mathcal{L}_{\text{mag-pha}} = \alpha\mathcal{L}_{\text{mag}} + \beta \mathcal{L}_{\text{pha}},
\end{equation}
where $\mathcal{L}_{\text{mag}}$ and $\mathcal{L}_{\text{pha}}$ denote the loss components associated with magnitude and phase, respectively.
The weighting coefficients $\alpha$ and $\beta$ serve to adjust the relative importance of these terms within the loss function.

The magnitude-related loss $\mathcal{L}_{\text{mag}}$ in Eq. (\ref{eq: magnitude-phase loss}) can take various forms, depending on whether a masking (\cite{phase_aware_xiaolei_zhang_TASLP_2019}) or mapping (\cite{PHASEN}) strategy is employed for magnitude spectrogram estimation.
Likewise, the phase-related loss $\mathcal{L}_{\text{pha}}$ differ across studies, including but not limited to the $L^p$ norm (\cite{ phase_aware_xiaolei_zhang_TASLP_2019, PHASEN, Training_strategy_masking+mapping_MP_SENet}), and cosine loss function (\cite{real_VAE_complex_spec_2019_ICASSP}).

\subsubsection{Time-Domain Loss Functions}
Since audio signals are ultimately perceived by humans in their raw waveform, time-domain loss has been chosen to better align with human auditory perception.
Time-domain penalties have been explored through diverse methodologies.
These approaches can broadly be classfied into two categories:
(1) computing direct distance measures between the estimated waveform and the ground truth, such as mean squared error (MSE) or mean absolute error (MAE);
or (2) utilizing evaluation metrics as loss functions to enhance model performance in terms of speech quality or intelligibility.
Several evaluation metrics have been used as loss functions, including the signal-to-distortion ratio (SDR) (\cite{SDR}), segmental signal-to-noise ratio (SSNR) (\cite{SSNR}), and scale-invariant signal-to-distortion ratio (SI-SDR) (\cite{SI-SDR}), with SI-SDR being the most prevalent in recent studies.
The SI-SDR loss is defined as:
\begin{equation}
\label{eq: SI-SDR}
    \mathcal{L}_{\text{SI-SDR}} = -10  \log_{10}
    \frac{\|\alpha \cdot \mathbf{x}\|^2}{\|\alpha \cdot \mathbf{x}-\hat{\mathbf{x}}\|^2},
\end{equation}
where,
\begin{equation}
    \alpha  = \frac{\hat{\mathbf{x}}^T \mathbf{x}}{ \|\mathbf{x}\|^2}.
\end{equation}
Here, $\mathbf{x}$ and $\hat{\mathbf{x}}$ denote the ground truth and estimated time-domain speech signal, respectively.


\subsubsection{Multi-Domain Loss Functions}
\label{sec: 5.2.4 multi-domain loss functions}
Despite the intuitive appeal of loss functions in the complex or time domains, experimental results have shown only limited performance improvements, particularly in terms of magnitude-based metrics (\cite{pandey2019exploring, compensation_IEEESignalLetters_Wang_Jonathan}).
Consequently, several studies have explored the design of loss functions with the aim of improving the model performance on evaluation metrics.
Speech signal evaluation typically takes multiple factors into account.
For instance, SSNR and SI-SDR measures are computed in the time domain, whereas perceptual evaluation of speech quality (PESQ), short-time objective intelligibility (STOI), extended STOI (ESTOI), and log-spectral distortion (LSD) primarily focus on the magnitude spectrograms. 
Since evaluation metrics account for multiple domains, a feasible approach to loss function design is to incorporate losses from multiple domains, ensuring a more comprehensive optimization strategy.

Given the critical importance of the magnitude spectrogram in a wide array of evaluation metrics, many studies have investigated the integration of penalties with magnitude and complex-domain losses.
For instance, \cite{fu_szu_wei_multi_task_complex_MLSP_2017} introduced a multi-metric loss function that incorporates the L2-norm of the complex spectrogram along with log-spectral distortion.
Despite this innovative approach, the reported performance improvements were modest, possibly due to an imbalance among the different terms within the composite loss function.

\cite{wangzhongqiu_loss_ICASSP_2020} further advanced the field by proposing a loss function that applies the L1 loss to the real, imaginary and magnitude spectrograms.
Its effectiveness has been validated in multiple domains, including speech resynthesis (\cite{CRVAE_yuying_xie}), speech enhancement (\cite{wangheming_enhancement_ICASSP_2022_bone_air}), and speaker separation (\cite{compensation_IEEESignalLetters_Wang_Jonathan}).
The promising performance improvements achieved by this multi-domain loss functions have spurred further research into loss function design.
Several subsequent studies have explored this topic, including theoretical analyses (\cite{compensation_IEEESignalLetters_Wang_Jonathan}) and experimental comparisons (\cite{braun2021consolidated}) to gain deeper insight into effectiveness.
For instance, \cite{compensation_IEEESignalLetters_Wang_Jonathan} provided a theoretical analysis of the performance improvement, framing it as a compensation problem.
Let $S(l,k)$ and $\hat{S}(l,k)$ denote the ground truth and the estimated spectrogram bin, respectively. 
In general, phase estimation is challenging, meaning that $\angle \hat{S}(l,k)$ can deviate significantly from $\angle {S}(l,k)$, especially when the input signal has a low SNR.
If the phase error $\angle S(l,k)-\angle  \hat{S}(l,k)$ exceeds $\pi / 2$, loss functions that consider only the direct distance $S(l,k)-\hat{S}(l,k)$ may drive the network to predict near-zero magnitudes, i.e. $|\hat{S}(l,k)|$ approaching zero, ultimately leading to ineffective predictions.
This explains why training solely with phase-aware losses alone does not yield strong performance on magnitude-sensitive metrics such as PESQ and ESTOI.
Incorporating a magnitude loss into the total loss function helps mitigate this issue by encouraging the DNN to achieve a balanced optimization in both complex and magnitude domains.
Moreover, \cite{braun2021consolidated} investigated the impact of various loss functions on evaluation performance in a lightweight RNN architecture for online speech enhancement.
Their study demonstrated that combining magnitude-only and phase-aware losses consistently led to performance improvements.
Additionally, they found that using compressed spectrals as features can further enhance performance, which is consistent with the results from \cite{Andong_Li_JASA_importance_of_compression}.
Subsequently, the design of loss functions has been extended to incorporate penalties in multiple domains.
For example, \cite{loss_triple_wang2021neural} proposed an objective function under a cascade architecture, which combines mask-based, complex-domain, and time-domain components.

Futhermore, most of the aforementioned methods compute the loss function using a single set of STFT parameters, thereby limiting the optimization of complex spectrogram to a single resolution. 
To overcome this limitation, \cite{loss_multi_resolution} recently proposed a multi-resolution loss function that combines losses computed across multiple STFT window lengths into a unified objective.
Experimental results have shown that incorporating penalties computed over multiple window lengths can further improve model performance.

\section{Applications}
\label{sec: 6. applications}

Complex spectrogram processing has been applied to a variety of speech tasks, including but not limited to phase retrieval, speech enhancement, speaker separation, emotion recognition (\cite{Cplx_emotion_recog}) and sound source localization (\cite{Cmplx_sound_localization}).
However, it should be noted that not all speech-related tasks require explicit phase modeling, since many of them can achieve satisfactory performance using magnitude features alone.
In speech signal processing, the primary benefits of incorporating phase information are often associated with improving perceptual quality for human listeners.

In this chapter, we focus on three major applications where DNN-based approaches have demonstrated substantial advancements in leveraging complex spectrograms: (1) phase retrieval (also known as phase reconstruction), which aims to estimate the phase spectrogram from a given magnitude spectrogram; (2) speech enhancement, focused on improving the clarity, intelligibility, and overall quality of speech signals in the presence of noise, reverberation, or other distortions; and (3) speaker separation, which seeks to isolate individual speech signals from mixtures containing overlapping voices.

\subsection{Phase Retrieval}
\label{sec: 6. phase retrieval}

Classical approaches to phase retrieval (\cite{gerkmann2015phase_overview}) can be broadly categorized into consistency-based methods, such as the well-known Griffin-Lim algorithm (GLA) (\cite{phase-retrieval-griffin-lim}), and model-based methods (\cite{phase-retrieval-model-based}).
With the advancements in deep learning and their remarkable ability to model complex partterns, many recent studies have explored leveraging DNNs to improve phase retrieval performance.

Nonetheless, direct phase retrieval using DNNs remains a challenging task.
One major obstacle is the phase wrapping issue.
As phase is naturally periodic, its value is only available between $[-\pi,\pi)$.
This nature not only causes discontinuity in training but also renders typical regression loss like MSE ineffecient.
Another significant challenge lies in phase sensitivity to waveform shifts.
Unlike the magnitude spectrogram, phase is highly sensitive to time shifts, with even minor shifts causing substantial variations. 
Moreover, additional complexities such as sign indertermination (\cite{phase_retrieval_sign_inderterminacy}) further complicate the phase retrieval process, making it a nontrival problem for deep learning-based approaches.

Given these challenges, researchers have developed alternative approaches. 
As phase-related loss functions were discussed in Subsection~\ref{sec: 5.2.1 phase-related loss functions}, below we will focus on deep learning-based phase retrieval methods.

One approach is to directly estimate the complex spectrogram, thus bypassing the need for direct phase estimation.
For instance, \cite{phase_retrieval_gan1} proposed using GAN to reconstruct the complex spectrogram from the known magnitude.
During training, the generator reconstructs the complex spectrogram, while the discriminator evaluates it against the ground truth.
\cite{phase_retrieval_anti_wrapping_zhenhua_ling} proposed a discriminative framework for phase retrieval via parallel spectrogram estimation. 
Their architecture employs a dual-branch mechanism that simultaneously estimates pseudo-real and pseudo-imaginary components to facilitate phase reconstruction.

Another approach is to integrate DNNs into classical approaches.
A case in point is the work by \cite{phase_retrieval_deep_griffin_lim_icassp}, who proposed the deep Griffin-Lim iteration (DeGLI).
This approach integrates a real-valued DNN-based sub-module within each GLA iteration cycle.
The integration of the DNN architecture serves to mitigate errors introduced by the GLA.

Additionally, several works have framed phase retrieval as a classification problem instead of a regression task.
The PhaseNet framework, introduced by \cite{phase_retrieval_classification_phasenet}, examplifies this approach by segmenting the phase into distinct classes with equidistant intervals and subsequently employing a DNN to predict the discrete phase.
Expanding on this concept, \cite{phase_retrieval_classification_phasebook} explored the use of an optimized phasebook, essentially a set of nonuniformly spaced classes, for phase discretization.

As mentioned in Section \ref{sec: 2. complex spectrogram representation}, phase derivatives (i.e., GD and IF) offer more structured representations than phase spectrograms. 
Early work (\cite{phase_retrieval_cosine_loss}) showed that estimating phase derivatives is more feasible than estimating phase directly. 
This led to the development of two-stage phase reconstruction approaches (\cite{phase_retrieval_two_stage1}), where the first stage uses DNNs to estimate phase derivatives, and the second reconstructs the phase spectrogram from them.
Subsequent research has expanded this approach, including: (1) exploring different DNN frameworks (\cite{phase_retrieval_two_stage2}), (2) alternative phase derivative representations like baseband phase delay (\cite{phase_retrieval_two_stage2}) and inter-frequency phase difference (\cite{phase_retrieval_two_stage4}), and (3) various phase reconstruction techniques, e.g. recurrent phase unwrapping (\cite{phase_retrieval_two_stage1}) and likelihood-based methods using the von Mises distribution (\cite{phase_retrieval_two_stage3, phase_retrieval_two_stage4}).

\subsection{Speech Enhancement}
\label{sec: 6.2 speech enhancement}
In this subsection, we review recent advancements in processing complex spectrograms for speech enhancement, especially focusing on discriminative models.
For generative model-based speech enhancement methods, please referred to Section~\ref{sec: 7. complex spectrograms and generative models}.

As the phase spectrogram lacks a clear structure, making it inherently difficult to predict directly using DNNs (\cite{cIRM_TASLP}), one research direction focusing on processing the real and imaginary spectrograms has emerged in speech enhancement.

The development of DNN-based complex spectrogram enhancement originated from real-valued DNN approaches.
Pioneering work by \cite{cIRM_ICASSP, SE_CIRM_williamson_2017taslp} proposed a real-valued DNN framework for estimating the complex ideal ratio mask (cIRM) in speech enhancement tasks.
Their experimental results demonstrated that cIRM estimation outperforms magnitude mapping and masking approaches in speech enhancement.
In parallel, \cite{fu_szu_wei_multi_task_complex_MLSP_2017} proposed using a real-valued CNN-based model for complex spectrogram mapping. 
The framework from \cite{fu_szu_wei_multi_task_complex_MLSP_2017} treats the real and imaginary components of noisy spectrograms as separate input channels, incorporating a multi-metric learning strategy, as discussed in Subsection~\ref{sec: 5.2.4 multi-domain loss functions}.

With the growing interest in CVNNs (\cite{deep_complex_networks}) and the fact that spectrograms are naturally complex-valued, several studies have extended DNNs from the real-valued domain to the complex domain.
For instance, \cite{complex_unet} proposed deep complex U-Net (DCUNET). 
Its components, including complex-valued convolutional layers, complex batch normalization and leaky $\mathbb{C}$ReLU, have been discussed in Sections~\ref{sec: 3. Key Components of Neural Networks} and~\ref{sec: 4. complex-valued neural networks}.
With incorporating a time-domain loss function during training, DCUNET was subsequently adopted for speech enhancement through complex mask estimation. 
It is worth noting that experimental results from \cite{complex_unet} reveal that, consistent with observations in RVNNs, a deeper complex-valued architecture exhibits superior enhancement performance.

While aforementioned CNN-based models have shown promising results, their limited receptive fields restrict them to local feature extraction. 
However, speech signals require modeling long-range dependencies across time and frequency. 
To overcome this, recent work has explicitly incorporated joint time-frequency context modeling.
\cite{SE_CRN_ke_tan_2019icassp, SE_CRN_ke_tan_2019taslp} introduced a real-valued convolutional recurrent network (CRN) specifically designed for complex spectrogram mapping.
Experimental results demonstrate the effectiveness of combining the strengths of convolutional operations for local feature extraction and recurrent structures for temporal modeling. 
Furthermore, \cite{hu2020dccrn} proposed a complex-valued architecture known as the Deep Complex Convolution Recurrent Network (DCCRN). 
By estimating a complex-valued mask and optimizing with an SI-SDR loss, DCCRN achieved superior performance over existing baselines while maintaining significantly lower computational complexity.

Subsequent research has focused on leveraging both cross-band and sub-band information to more effectively capture time–frequency dependencies.
This joint modeling strategy allows the network to exploit fine-grained spectral details within sub-bands while also accounting for global interactions across bands, which has been shown to enhance both objective metrics and perceptual quality in enhancement.
Representative real-valued models in this direction include FullSubNet (\cite{hao2021fullsubnet}), DPT-FSNet (\cite{SE_DPT-FSNet_ICASSP2022}), CMGAN (\cite{SE_generative_CMGAN}), and BSRNN (\cite{RI_concat_USR-BSRNN}). 
Similar ideas have also been explored in complex-valued models, such as DCCRN+ (\cite{SE_DCCRN+}).


Although modeling the phase spectrogram remains a significant challenge, recent studies have investigated speech enhancement through joint magnitude–phase processing (\cite{PHASEN, SE_DB_AIAT_ICASSP2022, SE_Uformer_ICASSP2022, Training_strategy_masking+mapping_MP_SENet, Training_strategy_masking+mapping_xLSTM_SENet}).
As discussed in Section~\ref{subsec: RVNN for magnitude-phase}, their architectures typically adopt either a dual-branch mechanism—where one branch estimates the magnitude spectrogram and the other focuses on phase or complex representations with cross-branch information exchange—or an encoder–decoder structure, and are therefore not elaborated here.
Building on the success of magnitude spectrogram processing, masking remains a central component for magnitude estimation in these works, while a parallel branch is often employed for phase mapping.

\subsection{Speaker Separation}
\label{sec: 6.3 speaker separation}
Since the introduction of deep cluster (\cite{SS_deep_cluster_2016icassp}) and permutation invariant training (PIT) (\cite{SS_PIT_2017ICASSP, mortern_PIT}), the performance of DNN-based speaker separation methods has seen substantial improvements.
Early approaches incorporated phase information using the phase sensitive mask (PSM) (\cite{erdogan2015PSM}), which yielded promising results in speaker separation (\cite{mortern_PIT}). 
However, PSM is applied only to the magnitude spectrogram, while the mixture phase is still used for inverse STFT, limiting its effectiveness. 
A more suitable approach involves estimating the real and imaginary spectrograms of separated sources.
In this direction, \cite{cIRM_TASLP} proposed using the complex-valued ideal ratio mask (cIRM) for speaker separation.
Both objective and subjective evaluations demonstrated the effectiveness of estimating phase spectrogram in improving separation quality.
Additionally, \cite{SS_l1_loss_2021TASLP} introduced a direct complex spectrogram mapping approach, training the model with a novel multi-domain loss.
Compared to conventional objective functions, this novel loss function led to significant performance improvements in speaker separation (\cite{compensation_IEEESignalLetters_Wang_Jonathan}).
Despite these advancements, most speaker separation research has focused on time-domain approaches, which require smaller window and hop sizes and have demonstrated superior performance.
Recently, \cite{SS_TF_GridNet_2023ICASSP} proposed the TF-GridNet framework, which leverages both global and local information through a multi-path architecture.
This model integrates intra-frame, sub-band, and full-band modules to enhance separation performance.
Experimental results demonstrate that TF-GridNet outperforms other state-of-the-art time-domain speaker separation baselines in terms of the SI-SDR metric.
Subsequently, SpatialNet (\cite{SS_quan2024spatialnet}) and TF-CrossNet (\cite{SS_kalkhorani2024tf_TF_CrossNet}) have been introduced, achieving superior results with reduced computational cost in speaker separation.

It is worth noting that this subsection offers only a concise overview of monaural speaker separation, with an emphasis on complex spectrogram processing. 
For a more comprehensive review of speaker separation techniques, readers are referred to the recent work by \cite{araki202530+_separation_overview} and \cite{SS_overview2_THU}.

\section{Generative Models for Complex Spectrograms}
\label{sec: 7. complex spectrograms and generative models}

Generative models are a class of machine learning models that learn the underlying data distribution from a given dataset, enabling them to generate new samples — such as images, text, or audio — that closely resemble real data.
Unlike traditional models that are primarily designed for classification or regression tasks, generative models excel at capturing complex patterns and variations, allowing for the synthesis of entirely novel and diverse content.
Recent breakthroughs in models like GANs (\cite{GAN_original_paper_2014NIPS}), VAEs (\cite{vae_original}), and diffusion models (\cite{denoise_diffusion}) have significantly advanced the field, leading to the generation of photorealistic images, high-fidelity audio, and even synthetic voices and videos.

In speech signal processing, most generative models operate on spectral representations, such as magnitude spectrograms or Mel-spectrograms, for synthesis tasks. 
Typically, a vocoder is employed to invert these spectral features back to waveform signals, as seen in models like WaveNet (\cite{van2016wavenet}) and HiFiGAN (\cite{kong2020hifigan}).
Despite significant advancements in vocoder technology, this cascaded structure often introduces distortions in the final generated audio. 
These distortions arise both from the vocoder itself and from the mismatch between the ideal training feature used for the vocoder and the feature generated by the generative model.
While the vocoder can be fine-tuned using the generated spectral feature, this process comes with increased computational costs.
Additionally, vocoders are typically constrained by fixed sampling frequency requirements, limiting their flexibility.
Recently, there has been growing interest in using generative models to directly handle complex spectrograms.
Below, we provide a brief summary of existing works, which could serve as a valuable foundation for future research.

The first part we introduce here is the VAE-related works.
The vanilla VAE architecture comprises two components: an inference network (encoder) and a generative network (decoder).
The encoder maps data $\mathbf{x}$ to a probabilistic latent space, while the decoder reconstructs $\mathbf{x}$ from samples in the latent space.
To handle complex spectrograms, various VAE variants have been proposed.
Broadly, these variants can be categorized into two major approaches. 
The first approach modifies only the decoder distribution while keeping the latent variable space as a real-valued Gaussian distribution.
In these works, the decoder may directly model the complex spectrogram as a whole (\cite{vae_nmf1_2018icassp_jp,vae_nmf4_apsipa2018_jp,vae_nmf5_mlsp2018_inria,vae_nmf2_2019icassp_inria_semi,vae_nmf3_dafx2019_inria_notes}), or model the magnitude and phase spectrograms separately (\cite{real_VAE_complex_spec_2019_ICASSP,real_vae_complex2_ucla_meta}).
Studies from (\cite{vae_nmf1_2018icassp_jp,vae_nmf4_apsipa2018_jp,vae_nmf5_mlsp2018_inria,vae_nmf2_2019icassp_inria_semi,vae_nmf3_dafx2019_inria_notes}) exemplify the first case, where the decoders in VAEs are modeled under the assumption of a zero-mean complex circularly symmetric Gaussian distribution.
However, relying solely on magnitude-based spectrograms as input introduces inherent limitations and inconsistencies in these works.
\cite{real_VAE_complex_spec_2019_ICASSP} and \cite{real_vae_complex2_ucla_meta} proposed approaches to model the magnitude and phase of spectrograms using distinct distribution assumptions.
Notwithstanding, the experimental results in both studies exhibit limited performance in phase reconstruction.

The second approach leverages VAEs for complex spectrogram modeling by extending the entire framework into the complex-valued domain, encompassing both the latent variables and the observations.
\cite{nakashika2020_CVAE} introduced an innovative complex-valued VAE (CVAE), in which both the observations and latent variables are assumed to follow a complex Gaussian distribution $\mathcal{N}_{C}(\mu, \sigma, \delta)$.
The assumptions enable the CVAE to not only process complex-valued data but also extend the latent variables to the complex domain.
Moreover, leveraging the inherent temporal dependencies in speech, \cite{CRVAE_yuying_xie} proposed a complex-valued recurrent VAE (CRVAE) that integrates complex-valued GRU layers, as discussed in Section~\ref{sec: 4. complex-valued neural networks}.
Inspired by the success of multi-domain loss functions, CRVAE adopts a deterministic decoder. 
Experimental results from both studies highlight the effectiveness of complex-valued VAEs in modeling complex spectrograms.


GANs (\cite{GAN_original_paper_2014NIPS}) represent a class of powerful generative models, composed of two neural networks — a generator and a discriminator — trained in a minimax game. 
While classical GANs learn to generate realistic data from random noise in an unsupervised manner (\cite{GAN_original_paper_2014NIPS}), recent applications in complex spectrogram processing typically adopt the framework of conditional GANs, where generation is guided by an additional input condition.
Applications of GANs on complex spectrograms processing are mostly focusing on speech enhancement, and also phase retrieval and audio synthesis.
In speech enhancement, the condition is the noisy speech signal itself, which implicitly contains both noise and the underlying clean speech. 
From a generative modeling perspective, the objective is to learn to conditionally generate clean complex spectrograms from noisy inputs. 
Specifically, the generator is trained to produce enhanced spectrograms using either real-valued or complex-valued neural networks (\cite{SE_generative_Skipconvgan_TASLP2022, Training_strategy_masking+mapping_MP_SENet, Training_strategy_masking+mapping_xLSTM_SENet, SE_generative_CMGAN}).
The discriminator, meanwhile, plays a critical role in shaping the learning dynamics by evaluating how realistic the generated spectrograms are. 
Beyond basic real-vs-fake classification (\cite{SE_generative_Skipconvgan_TASLP2022}), many studies have enhanced the discriminator with perceptual constraints by introducing feature-level distance metrics or differentiable perceptual evaluation scores, such as PESQ (\cite{SE_generative_CMGAN, SE_generative_CMGAN_extension, Training_strategy_masking+mapping_MP_SENet}). 
These additions enable the model to better align with human auditory perception, reinforcing the idea of GANs as perceptually-aware generative models.

Beyond speech enhancement, GANs have also been applied to other generation tasks, such as phase retrieval (\cite{phase_retrieval_gan1}) and audio synthesis. 
A notable example of the latter is GANSynth (\cite{GANSynth}), which demonstrates that modeling log magnitudes and instantaneous frequencies in the complex spectrogram domain, with sufficient frequency resolution, enables GANs to generate high-quality, locally coherent audio.


Diffusion models (\cite{SE_generative_diffusion_continuous_SDE}) are generative models that learn to gradually denoise data by reversing a fixed forward noising process, typically formulated as a Markov chain or a continuous stochastic differential equation (SDE).

The works of applying diffusion models to complex spectrograms are predominantly in speech enhancement. 
Similar with the cases in GAN, diffusion models can be regarded as conditional generative models in such scenarios.
Building upon the foundation established by \cite{SE_generative_diffusion_continuous_SDE}, \cite{SE_generative_diffusion_Gerkmann_2022InterSpeech} introduced a continuous SDE framework for speech enhancement, named SGMSE. 
This generalized continuous formulation enables flexible integration of arbitrary numerical solvers during the reverse diffusion process.
As the model is proposed for complex spectrograms, the perturbation kernel used in SGMSE is a multivariant circularly-symmetric complex Gaussian distribution $\mathcal{N}_{C}(\mu,\sigma)$.
Moreover, advancements of diffusion models have been further used to handle complex spectrogram-based enhancement, including cold diffusion (\cite{SE_generative_diffusion_cold_2023_MERL}), unconditional diffusion model (\cite{SE_generative_diffusion_with_unconditional}).
Recently, \cite{Philippe_diffusion_TASLP} extends a diffusion model originally proposed for image generation to incorporate a speech enhancement system, where the diffusion process has a non-zero long-term mean equal to the conditioning signal. 
This approach directly models the clean speech signal and offers a more general formulation.
Complex-valued backbones like DCUNET (\cite{complex_unet}) and DCCRN (\cite{hu2020dccrn}), real-valued backbones include Noise Conditional Score Network (NCSN++) (\cite{SE_generative_diffusion_continuous_SDE}), CRN (\cite{SE_CRN_ke_tan_2019taslp}) have been explored for diffusion model-based complex spectrogram processing across works (\cite{SE_generative_diffusion_Gerkmann_2022InterSpeech, SE_generative_diffusion_cold_2023_MERL, SE_generative_diffusion_Gerkmann_2023TASLP, SE_generative_diffusion_with_discriminative1}).

Additionally, diffusion models are often combined with discriminative models in a hybrid framework.
The discriminative model first provides an initial estimate of clean speech, which is then refined by the diffusion model to further reduce distortions using techniques such as denoising diffusion restoration and stochastic regeneration (\cite{SE_generative_diffusion_with_discriminative1, SE_generative_diffusion_with_discriminative2, SE_generative_diffusion_with_discriminative3, SE_generative_diffusion_with_discriminative4}). 
This hybrid method avoids generating artifacts such as vocalizing and breathing effects while significantly reducing computational costs (\cite{SE_generative_diffusion_with_discriminative4}).
More recently, low-latency diffusion model-based speech enhancement has been explored (\cite{latency_timo_lay2025diffusion}).
For a more comprehensive exploration of this topic, \cite{diffusion_overview} offered a detailed analysis in their work.

\section{Conclusion}
\label{sec: 10. conclusion}
This survey has comprehensively explored the advancements and methodologies in processing complex spectrograms using DNNs. 
We delved into the fundamental aspects of complex spectrograms and the key components of CVNNs. 
The approaches of using RVNNs for handling complex spectrograms have also been summarized.
Various training strategies and loss functions were discussed for learning complex spectrograms. 
Additionally, we examined the DNN-based applications in phase retrieval, speech enhancement, and speaker separation, showcasing their effectiveness in improving audio signal quality. 
Finally, the intersection of complex spectrograms with generative models was explored.

\bibliographystyle{elsarticle-harv} 
\bibliography{example}





\end{document}